\documentclass[journal]{IEEEtran}

\usepackage{url}
\usepackage{cite}
\usepackage{array}
\usepackage{amsthm}
\usepackage{dsfont}
\usepackage{amsmath}
\usepackage{amsfonts}
\usepackage{booktabs}
\usepackage{graphicx}
\usepackage{multicol}
\usepackage{multirow}
\usepackage{stfloats}
\usepackage{textcomp}
\usepackage{verbatim}
\usepackage{cellspace}
\usepackage{enumerate}
\usepackage{enumitem}
\usepackage[T1]{fontenc}
\usepackage[dvipsnames]{xcolor}
\usepackage[ruled,vlined,linesnumbered]{algorithm2e}
\usepackage[caption=false,font=normalsize,labelfont=sf,textfont=sf]{subfig}

\theoremstyle{plain}
\newtheorem{definition}{Definition}

\newtheorem{theorem}{Theorem}

\newcommand{\header}[1]{\vspace{0.5mm}\noindent \textbf{#1}}

\newcommand{\tabincell}[2]{\begin{tabular}{@{}#1@{}}#2\end{tabular}}

\def\eps{\varepsilon}

\def\tO{\widetilde{O}}

\def\A{\mathbf{A}}
\def\D{\mathbf{D}}
\def\I{\mathbf{I}}
\def\P{\mathbf{P}}
\def\bPi{\mathbf{\Pi}}

\def\Nin{\mathcal{N}_{\mathrm{in}}}
\def\Nout{\mathcal{N}_{\mathrm{out}}}
\def\din{d_{\mathrm{in}}}
\def\dout{d_{\mathrm{out}}}

\def\vec{\boldsymbol}
\def\ex{\hat{x}}
\def\vx{\vec{x}}
\def\vex{\vec{\ex}}
\def\vpi{\vec{\pi}}
\def\vpiinv{\vpi^{-1}}
\def\vsigma{\vec{\sigma}}
\def\vepi{\vec{\hat{\pi}}}

\def\vepif{\vepi^{(\mathrm{f})}}
\def\vepib{\vepi^{(\mathrm{b})}}
\def\ve{\vec{e}}
\def\vzero{\vec{0}}
\def\vone{\vec{1}}
\def\vr{\vec{r}}
\def\vrf{\vr^{(\mathrm{f})}}
\def\vrb{\vr^{(\mathrm{b})}}

\DeclareMathOperator{\E}{E}

\def\rmax{r_\mathrm{max}}
\def\rmaxf{\rmax^{(\mathrm{f})}}
\def\rmaxb{\rmax^{(\mathrm{b})}}

\def\Alg{Alg.}
\def\Eqn{Eqn.}
\def\Eqns{Eqns.}
\def\Fig{Fig.}
\def\Sec{Sec.}
\def\Tbl{Tbl.}
\def\Tbls{Tbls.}

\def\Thm{Thm.}

\def\jump{\textsc{jump}}

\begin{document}

\title{Efficient Algorithms for Personalized PageRank Computation: A Survey}

\author{Mingji~Yang, Hanzhi~Wang, Zhewei~Wei, Sibo~Wang, and Ji-Rong~Wen
    \IEEEcompsocitemizethanks{
    \IEEEcompsocthanksitem Zhewei Wei is the corresponding author. 
    \IEEEcompsocthanksitem Mingji Yang, Hanzhi Wang, Zhewei Wei, and Ji-Rong Wen are with Renmin University of China.
    Zhewei Wei is with Gaoling School of Artificial Intelligence, Beijing Key Laboratory of Big Data Management and Analysis Methods, MOE Key Lab of Data Engineering and Knowledge Engineering, and Pazhou Laboratory (Huangpu), Guangzhou, Guangdong 510555, China.
    Emails: \{kyleyoung, hanzhi\_wang, zhewei, jrwen\}@ruc.edu.cn
    \IEEEcompsocthanksitem Sibo Wang is with The Chinese University of Hong Kong.
    Email: swang@se.cuhk.edu.hk
    }%
    \thanks{\copyright~2024 IEEE. Personal use of this material is permitted. Permission from IEEE must be obtained for all other uses, in any current or future media, including reprinting/republishing this material for advertising or promotional purposes, creating new collective works, for resale or redistribution to servers or lists, or reuse of any copyrighted component of this work in other works.}
}

\markboth{This text is the ``accepted version'' of an article accepted by IEEE TKDE for publication.}%
{}

\maketitle

\begin{abstract}
Personalized PageRank (PPR) is a traditional measure for node proximity on large graphs.
For a pair of nodes $\boldsymbol{s}$ and $\boldsymbol{t}$, the PPR value $\boldsymbol{\vpi_s(t)}$ equals the probability that an $\boldsymbol{\alpha}$-discounted random walk from $\boldsymbol{s}$ terminates at $\boldsymbol{t}$ and reflects the importance between $\boldsymbol{s}$ and $\boldsymbol{t}$ in a bidirectional way.
As a generalization of Google's celebrated PageRank centrality, PPR has been extensively studied and has found multifaceted applications in many fields, such as network analysis, graph mining, and graph machine learning.
Despite numerous studies devoted to PPR over the decades, efficient computation of PPR remains a challenging problem, and there is a dearth of systematic summaries and comparisons of existing algorithms.
In this paper, we recap several frequently used techniques for PPR computation and conduct a comprehensive survey of various recent PPR algorithms from an algorithmic perspective.
We classify these approaches based on the types of queries they address and review their methodologies and contributions.
We also discuss some representative algorithms for computing PPR on dynamic graphs and in parallel or distributed environments.
\end{abstract}


\section{Introduction} \label{sec:intro}

\IEEEPARstart{G}{raphs} are a ubiquitous abstract data type that models a wide range of information in the real world.
\textit{Personalized PageRank (PPR)}~\cite{brin1998anatomy}, as a direct variant of Google's renowned \textit{PageRank} centrality~\cite{brin1998anatomy}, is a widely used measure of node proximity in graphs.
Given a source node $s$ and a target node $t$, the PPR value $\vpi_s(t)$ offers a quantitative amount of the significance of $t$ w.r.t. $s$.
On the other hand, $\vpi_s(t)$ also indicates the importance of $s$ in the view of $t$~\cite{wang2018efficient}, and thus it essentially gauges whether $s$ and $t$ are mutually significant.
The PageRank centrality of $t$, denoted as $\vpi(t)$, equals the average of $\vpi_s(t)$ over all source nodes $s$ and can be viewed as a special form of generalized PPR.

Although the seminal paper of PageRank and PPR~\cite{brin1998anatomy} considers utilizing them to rate webpages objectively and mechanically, recent years have witnessed a wide range of their applications beyond the web.
In fact, due to its simplicity and profound influence, PageRank was named one of the top-$10$ algorithms in data mining in 2008~\cite{wu2008top}.
We provide three representative instances of PPR's applications below: \textit{local clustering}, \textit{node embedding}, and \textit{graph neural networks}.
More applications can be found in Gleich's survey~\cite{gleich2015pagerank}.

\header{Local Graph Clustering.}
As a traditional graph mining task, local graph clustering aims at identifying a well-connected cluster around a specified node by only examining a local region of the graph.
A notable algorithm is PageRank-Nibble by Anderson, Chung, and Lang~\cite{andersen2006local,andersen2007pagerank}, which leverages a local-push method for computing PPR to find the local cluster efficiently with strong theoretical guarantees.
Following PageRank-Nibble, there arises a collection of recent works that rely on PPR or its extensions to compute local clusters, such as MAPPR~\cite{yin2017local}, TEA~\cite{yang2019efficient}, and \cite{fountoulakis2019variational,chang2022community,yuan2024index}.

\header{Node Embedding.}
Node embedding methods endeavor to learn low-dimensional representations of the nodes in the graph while preserving the similarities between them.
These methods typically require a node proximity measure in the first place, and many of them choose PPR values, such as HOPE~\cite{ou2016asymmetric}, APP~\cite{zhou2017scalable}, and VERSE~\cite{tsitsulin2018verse}.
Some recent works~\cite{yin2019scalable,yang2020homegeneous,tsitsulin2021frede} also leverage PPR or its extensions.

\header{Graph Neural Networks.}
Recently, Graph Neural Networks (GNNs) have fueled an explosion of research in graph machine learning.
As a fundamental measure of node proximity, PPR has played an important role in designing various GNNs, especially scalable ones.
For example, APPNP~\cite{klicpera2018predict} and PPRGo~\cite{bojchevski2020scaling} apply approximate PPR to GNNs, making them more efficient and scalable; GBP~\cite{chen2020scalable} and AGP~\cite{wang2021approximate} generalize the notion of PPR to propose novel methods for effective approximate graph propagation.

\IEEEpubidadjcol

Other recent extensions and applications of PPR include \cite{zhang2016accuracy,epasto2022differentially,yang2022efficient,zhang2023effective}.
Despite the abundant applications, the computation of PageRank and PPR remains a bottleneck in some situations~\cite{chen2020scalable,wang2021approximate}, as the graphs under consideration are becoming increasingly massive.
Even in 2002, the graph could contain $2.7$ billion nodes, and computing PageRank was consequently dubbed ``The World's Largest Matrix Computation''~\cite{moler2002world}.
Also, the complexity of computing PageRank and PPR is of individual theoretical interest.
These factors have sparked a large corpus of studies concerning the efficient computation of PPR on large graphs in the past decades.

Based on these studies, several surveys on PPR algorithms have been conducted, such as Berkhin's early survey in 2005~\cite{berkhin2005survey}, Chung's brief survey in 2014~\cite{chung2014brief}, and a more recent one by Park et al. in 2019~\cite{park2019survey}.
Nevertheless, none of them provides meticulous summaries of the fundamental techniques for computing PPR, nor do they review recent PPR algorithms comprehensively from a theoretical perspective.
In contrast, this paper strives to address these gaps by delving into five atomic techniques and systematically categorizing numerous algorithms according to their query types.
We provide theoretical analyses of the basic techniques, point out their pros and cons, and summarize the surveyed algorithms' methodologies and contributions.
Additionally, to cater to the emerging needs of handling evolving graphs, exploiting new computing environments, etc., we include a section for recent PPR algorithms tailored to these special settings.
While we primarily focus on the proximity measure of PPR, we also include numerous algorithms for PageRank computation.

\header{Paper Organization.}
The remainder of this paper is organized as follows.
\Sec~\ref{sec:pre} provides preliminary knowledge about PPR and defines various PPR computation problems.
\Sec~\ref{sec:basic_techniques} reviews five fundamental techniques extensively employed in PPR computation.
\Sec~\ref{sec:overview} gives an overview of various PPR algorithms.
In \Sec~\ref{sec:SSPPR_algorithms} to \ref{sec:other_queries}, we summarize the salient ideas and contributions of numerous recent PPR algorithms under the typical setting, categorized by the query types.
As a complement, \Sec~\ref{sec:special} discusses the algorithms on dynamic graphs, in parallel/distributed environments, or for other special settings.
After that, we discuss the research challenges and directions in \Sec~\ref{sec:directions}.
Finally, \Sec~\ref{sec:conclusion} concludes the paper.

\section{Preliminaries} \label{sec:pre}

This section offers preliminary knowledge for PPR computation, including basic notations for graphs, definitions of PPR and PPR queries, and some fundamental properties of PPR.

\subsection{Notations for Graphs} \label{sec:notations_graphs}

PageRank and PPR were originally proposed by Google as a tool to assign an authority score to each webpage solely based on the links between them~\cite{brin1998anatomy}.
Nonetheless, the notion of PageRank and PPR can be directly applied to any graph for measuring the importance of the nodes in the graph.
In light of this, we begin with some basic definitions of graphs.

We consider a directed or undirected graph $G=(V,E)$ with $|V|=n$ nodes and $|E|=m$ edges, and we assume that the nodes in $V$ are numbered in some order as $v_1,v_2,\ldots,v_n$.
Several fundamental concepts on $G$ are listed below.

\header{Neighbors and Degrees.}
For a node $v\in V$, we denote the set of the in-neighbors of $v$ by $\Nin(v)=\big\{u\mid(u,v)\in E\big\}$, and let $\din(v)=\big\lvert\Nin(v)\big\rvert$.
Similarly, we define $\Nout(v)=\big\{u\mid(v,u)\in E\big\}$ and $\dout(v)=\big|\Nout(v)\big|$.
For undirected graphs, we use $d(v)$ to denote the degree of $v$.

\header{Matrices of $\boldsymbol{G}$.}
The adjacency matrix of $G$ is an $n\times n$ matrix $\A$ whose $(i,j)$-th entry equals $\mathds{1}\big\{(v_i,v_j)\in E\big\}$, where $\mathds{1}\{\}$ is the indicator function.
The degree matrix of $G$ is an $n\times n$ diagonal matrix $\D$ whose $i$-th diagonal entry equals $\dout(v_i)$.
The transition matrix of $G$ is formalized as $\P=\A^{\top}\D^{-1}$.
To ensure that $\P$ is well-defined, we assume that $\dout(v)>0$ for all $v\in V$ throughout this paper (see the last paragraph in \Sec~\ref{sec:def_PPR} for discussions).
In this case, $\P$ is column-stochastic and can thus be viewed as a Markov chain transition matrix.

\header{Indicator Vector.}
For a node $v\in V$, we define $\ve_v$ as an $n$-dimensional column vector whose $i$-th entry equals $\mathds{1}\{v_i=v\}$.
Note that we use column vectors instead of row vectors throughout this survey.
Also, for an $n$-dimensional vector $\vx$ associated with the nodes in $V$, we denote by $\vx(v)$ the $i$-th entry of $\vx$, where $i$ satisfies $v_i=v$.

\subsection{Definitions of PPR and PageRank} \label{sec:def_PPR}

The concepts of PPR and PageRank are similar, and they can both be formulated as random-walk probabilities.
While this paper mainly concentrates on PPR computation, we also consider PageRank and regard it as a special form of generalized PPR.
We begin with the definition of PPR and then define PageRank based on it.

We first give the algebraic definition of PPR.
We denote the PPR vector w.r.t. a source node $s$ as $\vpi_s$, which assigns a proximity score $\vpi_s(t)$ to each $t\in V$.
$\vpi_s$ is defined as the unique solution to the linear equation
\begin{align}
    \vpi_s&=(1-\alpha)\P\vpi_s+\alpha\cdot\ve_s, \label{eqn:PPR_def}
\end{align}
where $\alpha\in(0,1)$ is a predefined parameter called the \textit{teleportation probability} (a.k.a. \textit{decay factor}).
Here, the indicator vector $\ve_s$ is called the \textit{preference vector} in defining PPR.
To demonstrate that $\vpi_s$ is well-defined, note that the matrix $\I-(1-\alpha)\P$ is guaranteed to be nonsingular since it is \textit{strictly column diagonally dominant}.
Thus, $\vpi_s$ equals
\begin{align}
    \vpi_s&=\alpha\big(\I-(1-\alpha)\P\big)^{-1}\ve_s, \label{eqn:PPR_inverse}
\end{align}
which establishes the existence and uniqueness of $\vpi_s$.
A more general definition of PPR allows the preference vector in \Eqn~\eqref{eqn:PPR_def} to be an arbitrary probability distribution vector, rather than being limited to indicator vectors.
As we will see from \Thm~\ref{thm:linearity}, these generalized PPR vectors are linear combinations of the PPR vectors in our definition.

\header{Random-Walk-Based Interpretations.}
Apart from the algebraic definition, there are two more comprehensible interpretations of PPR.
Both of them are based on a critical concept called \textit{random walk}, with two slightly different definitions.

In the first possible definition, a random walk sets out from the source node $s$, and at each step, it either (i) jumps back to $s$ w.p. (shorthand for ``with probability'') $\alpha$, or (ii) proceeds uniformly at random to an out-neighbor of the current node.
This type of random walk is also called \textit{random walk with restart (RWR)}.
The PPR vector $\vpi_s$ is then defined as the \textit{stationary distribution} over $V$ for an RWR after infinitely many steps.
For this reason, PPR values are occasionally referred to as \textit{RWR values}.

Alternatively, one can modify the procedure above to define an \textit{$\alpha$-discounted random walk}, which starts at $s$ and terminates at the current node (instead of jumping back to $s$) w.p. $\alpha$ at each step.
The vector $\vpi_s$ can then be defined as the \textit{termination distribution} over the nodes at which such an $\alpha$-discounted random walk eventually terminates.
It can be verified that the three definitions of PPR above are essentially equivalent~\cite{avrachenkov2007monte}.

Since the PPR value $\vpi_s(t)$ serves as a bidirectional proximity measure between nodes $s$ and $t$~\cite{wang2018efficient}, the \textit{inverse PPR vector} $\vpiinv_t$ is also of research interest, where $\vpiinv_t(s)$ is defined to be $\vpi_s(t)$.
For convenience, we further define the \textit{PPR matrix} $\bPi$ as an $n\times n$ matrix, with its $(i,j)$-th entry equaling $\vpi_{v_j}(v_i)$.
$\bPi$ encompasses $n^2$ possible PPR values between all nodes in $V$, while its $i$-th column equals $\vpi_{v_i}$ and $i$-th row equals $\vpiinv_{v_i}$.

\header{PageRank.}
We denote the PageRank vector as $\vpi$, which assigns a centrality score $\vpi(t)$ to each $t\in V$.
Algebraically, PageRank is a special form of generalized PPR with $1/n\cdot\vone$ as the preference vector, where $\vone$ is the $n$-dimensional vector with value $1$ on each coordinate:
\begin{align}
    \vpi&=(1-\alpha)\P\vpi+\alpha\cdot\frac{1}{n}\vone. \label{eqn:PR_def}
\end{align}
As we will see from \Thm~\ref{thm:linearity}, $\vpi$ equals the average of $\vpi_s$ over all $s\in V$.
PageRank can also be defined using RWR or $\alpha$-discounted random walks.
For this purpose, an RWR sets out from an arbitrary node and jumps back to a uniformly random node in $V$ w.p. $\alpha$ at each step, and an $\alpha$-discounted random walk starts from a uniformly random source node in $V$.

\header{Eliminating Dangling Nodes.}
We call a node $v$ a \textit{dangling node} if $\dout(v)=0$.
As mentioned, the formal definition of PPR and PageRank requires no dangling nodes in the graph.
However, this is hardly the case in practice.
When working with real-world graphs, several strategies for handling dangling nodes have been proposed.
Three commonly used remedies are: ({\romannumeral 1}) introduce a dummy node that has a self-loop to itself and add an edge from each dangling node to it; ({\romannumeral 2}) assume that each dangling node has outgoing edges to each node in $V$; and ({\romannumeral 3}) add a self-loop to each dangling node.
Different methods may incur minor differences in the resulting PPR values, but the overall ranking remains the same.
See, e.g., \cite{bianchini2005inside} for more detailed descriptions and analyses.

\subsection{Definitions of PPR Queries} \label{sec:def_PPR_queries}

As introduced above, the $n\times n$ PPR matrix $\bPi$ contains the PPR values between all nodes in $V$ as useful proximity scores.
However, computing and storing the dense matrix $\bPi$ requires $\Omega\left(n^2\right)$ time and space, which is prohibitive when $n$ reaches, say, $10^7$.
Fortunately, in many scenarios, it is unnecessary to compute the entire $\bPi$; instead, we only need a single row/column of $\bPi$ or even a single entry in it.
This motivates us to define various \textit{query types} for PPR computation.

\header{Query Types.}
Common queries for PPR computation include:
\begin{itemize}[leftmargin=*]
    \item \textit{single-source PPR (SSPPR)}: given a source node $s\in V$, the PPR vector $\vpi_s$ is required.
    \item \textit{single-target PPR (STPPR)}: given a target node $t\in V$, the inverse PPR vector $\vpiinv_t$ is required.
    \item \textit{single-pair PPR (SPPPR)}: given a source node $s\in V$ and a target node $t\in V$, the PPR value $\vpi_s(t)$ is required.
    \item \textit{top-$k$ PPR}: given a source node $s$ and an integer $k$, the nodes with the top-$k$ largest values in $\vpi_s$ are required.
    In some cases, their corresponding values in $\vpi_s$ are also required.
\end{itemize}
Additionally, for PageRank computation, two common tasks are the \textit{PageRank vector query}, which asks for the vector $\vpi$, and the \textit{single-node PageRank (SNPR) query}, which asks for the value $\vpi(t)$ of a single target node $t$.
They can be viewed as generalizations of the SSPPR and SPPPR queries, resp.
The whole PPR matrix can be obtained by invoking the SSPPR queries for all $s\in V$ or the STPPR queries for all $t\in V$.

As we will see, these queries can be answered \textit{exactly} using algorithms based on algebraic techniques.
However, computing exact answers often incurs significant costs, which is impractical for efficient online queries.
Particularly, for the SPPPR and top-$k$ PPR queries that have small output sizes, computing exact answers is unnecessarily wasteful.
In real applications, it is generally acceptable that these queries are answered \textit{approximately} in exchange for higher efficiency.
Also, introducing approximation allows algorithms to answer certain queries with sparse vectors.
The errors incurred by the approximate answers are typically quantified by the following four types of \textit{error measures}.

\header{Error Measures.}
For two $n$-dimensional vectors $\vex$ and $\vx$ or a positive real number $x$ along with its estimate $\hat{x}$, four commonly used error measures are defined as follows:
\begin{itemize}[leftmargin=*]
    \item \textit{$\ell_1$-error}: $\lVert\vex-\vx\rVert_1=\sum_{v\in V}\big\lvert\vex(v)-\vx(v)\big\rvert.$
    \item \textit{$\ell_2$-error}: $\lVert\vex-\vx\rVert_2=\sqrt{\sum_{v\in V}\big(\vex(v)-\vx(v)\big)^2}.$
    \item \textit{absolute error} (a.k.a. \textit{additive error} or \textit{$\ell_{\infty}$-error} for vectors): $\lVert\vex-\vx\rVert_{\infty}=\max_{v\in V}\big\lvert\vex(v)-\vx(v)\big\rvert$, or $|\ex-x|$.
    \item \textit{relative error with threshold $\mu$}: \\ $\max_{v\in V:\vx(v)\ge\mu}\frac{|\vex(v)-\vx(v)|}{\vx(v)}$, or $\frac{\lvert\ex-x\rvert}{x}\cdot\mathds{1}\{x\ge\mu\}$.
\end{itemize}
These definitions are straightforward, except for relative error: given the inherent difficulty of achieving the same relative error for smaller values, many relevant algorithms offer relative error bounds only for values above a threshold $\mu$.
Also, note that these error measures are not directly applicable for the top-$k$ PPR query; we will discuss the approximation guarantees separately for the top-$k$ PPR query.

\header{Probabilistic Algorithms.}
Furthermore, to achieve higher efficiency, it is common to allow the algorithms to be \textit{probabilistic}, as long as they can answer the queries successfully with high probability.
Concretely, these probabilistic algorithms receive an extra parameter $\delta$ as input and guarantee to return acceptable results w.p. at least $(1-\delta)$.
With approximation and randomness, the algorithms can potentially answer the queries in \textit{sublinear} time w.r.t. the size of the graph.

Combining the three factors above, a concrete algorithm is required to answer a specific type of PPR query exactly or under a designated error measure, and it may be allowed a small failure probability $\delta$.
We denote the error bound by $\eps$ and require that the designated error of the results does not exceed $\eps$.
Below, we define a specific query as an example.
\begin{definition}[Probabilistic SSPPR Query with Relative Error Bounds]
    Given a graph $G$, a decay factor $\alpha$, a source node $s$, a relative error parameter $\eps$, a relative error threshold $\mu$, and a failure probability parameter $\delta$, the query requires an estimated PPR vector $\vepi_{s}$ such that w.p. at least $(1-\delta)$, $\big\lvert\vepi_{s}(v)-\vpi_{s}(v)\big\rvert\le\eps\cdot\vpi_{s}(v)$ holds for all $v$ with $\vpi_{s}(v)\ge\mu$.
\end{definition}

\header{Conversions between Error Measures.}
Despite the differences between the error measures, it is possible to directly convert some error guarantees to another type using the following facts.
Here, we use $\eps_{\ell}$, $\eps_a$, and $\eps_r$ to denote the $\ell_1$-error, the absolute error, and the relative error, resp.
\begin{itemize}[leftmargin=*]
    \item from $\ell_1$-error to absolute error: $\eps_{\ell}\le\eps$ implies $\eps_a\le\eps$.
    \item from absolute error to $\ell_1$-error: $\eps_a\le\eps/n$ implies $\eps_{\ell}\le\eps$.
    \item from absolute error to relative error: $\eps_a\le\eps\mu$ implies $\eps_r\le\eps$ (with relative error threshold $\mu$).
\end{itemize}
However, these conversions are wasteful in general as they do not explicitly consider the features of the target error measures.

\begin{table}[!t]
\caption{Summary of Notations} \label{tab:notations}
\centering
    \begin{tabular}{SlSl}
        \toprule
        \textbf{Notation} & \textbf{Description} \\
        \midrule
    	$G=(V,E)$ & graph with node set $V$ and edge set $E$ (\Sec~\ref{sec:notations_graphs}) \\
    	$n,m$ & number of nodes/edges in $G$ (\Sec~\ref{sec:notations_graphs}) \\
    	$\Nin(v),\Nout(v)$ & the set of in/out-neighbors of node $v$ (\Sec~\ref{sec:notations_graphs}) \\
    	$\din(v),\dout(v)$ & in/out-degree of node $v$ (\Sec~\ref{sec:notations_graphs}) \\
        $\A,\D$ & adjacency and degree matrices of $G$ (\Sec~\ref{sec:notations_graphs}) \\
        $\P$ & transition matrix of $G$, $\P=\A^{\top}\D^{-1}$ (\Sec~\ref{sec:notations_graphs}) \\
        $\ve_s$ & indicator vector (\Sec~\ref{sec:notations_graphs}) \\
    	$\alpha$ & teleportation probability (\Sec~\ref{sec:def_PPR}) \\
    	$\vpi_s$ & PPR vector of source node $s$ (\Sec~\ref{sec:def_PPR}) \\
    	$\vpi$ & PageRank vector of $G$ (\Sec~\ref{sec:def_PPR}) \\
    	$\vpiinv_t$ & inverse PPR vector of target node $t$ (\Sec~\ref{sec:def_PPR}) \\
    	$\bPi$ & PPR matrix (\Sec~\ref{sec:def_PPR}) \\
    	$\eps$ & error bound parameter (\Sec~\ref{sec:def_PPR_queries}) \\
    	$\mu$ & relative error threshold (\Sec~\ref{sec:def_PPR_queries}) \\
    	$\delta$ & failure probability parameter (\Sec~\ref{sec:def_PPR_queries}) \\
    	$\rmaxf,\rmaxb$ & threshold of FP and BP, resp. (\Sec~\ref{sec:basic_techniques}) \\
    	$\vepif_s,\vepib_t$ & reserve vector in FP and BP, resp. (\Sec~\ref{sec:basic_techniques}) \\
    	$\vrf_s,\vrb_t$ & residue vector in FP and BP, resp. (\Sec~\ref{sec:basic_techniques}) \\
    	\bottomrule
    \end{tabular}
\end{table}

\subsection{Basic Properties of PPR} \label{sec:properties}

In this subsection, we overview several important properties of PPR without proving them.
These properties will be frequently used when designing algorithms for computing PPR.

Algebraically, the PPR vector $\vpi_s$ can be expressed as the following infinite summation:
\begin{align}
    \vpi_s=\alpha\sum_{\ell=0}^{\infty}(1-\alpha)^{\ell}\P^{\ell}\ve_s. \label{eqn:PPR_series}
\end{align}
Here, $\I$ is the identity matrix of order $n$.
The inverse PPR vector $\vpiinv_t$ satisfies $\vpiinv_t=(1-\alpha)\P^{\top}\vpiinv_t+\alpha\ve_t$ and equals
\begin{align}
    \alpha\left(\I-(1-\alpha)\P^{\top}\right)^{-1}\ve_t=\alpha\sum_{\ell=0}^{\infty}(1-\alpha)^{\ell}\left(\P^{\top}\right)^{\ell}\ve_t. \label{eqn:STPPR_series}
\end{align}
The PPR matrix $\bPi$ satisfies $\bPi=(1-\alpha)\P\bPi+\alpha\I$ and
\begin{align*}
    \bPi=\alpha\big(\I-(1-\alpha)\P\big)^{-1}=\alpha\sum_{\ell=0}^{\infty}(1-\alpha)^{\ell}\P^{\ell}.
\end{align*}
Some other algebraic properties of PPR are listed below, wherein $\lVert\vx\rVert_1$ denotes the $\ell_1$-norm of vector $\vx$.
\begin{align}
    \nonumber \lVert\vpi_s\rVert_1=1&,\ \left\lVert\vpiinv_t\right\rVert_1=n\vpi(t). \\
    \nonumber \vpi_s=\bPi\ve_s&,\ \vpiinv_t=\bPi^{\top}\ve_t. \\
    \nonumber \lVert\bPi\vx\rVert_1=\lVert\vx\rVert_1&,\ \bPi^{\top}\vone=\vone.
\end{align}

The following theorem considers the generalized PPR vector $\vpi_{\vsigma}$ for an arbitrary preference vector $\vsigma$.
It states that $\vpi_{\vsigma}$ is a linear combination of the PPR vectors $\vpi_s$.
\begin{theorem}[The Linearity Theorem~\cite{jeh2003scaling}] \label{thm:linearity}
    For a preference vector $\vsigma$, we have $\vpi_{\vsigma}=\sum_{s\in V}\vsigma(s)\cdot\vpi_{s}$.
\end{theorem}

The following theorem provides a method to decompose a node's PPR vector or inverse PPR vector into a linear combination of the corresponding vectors of its neighbors.
Intuitively, these formulas can be obtained by enumerating the first/last step of the random walk.
\begin{theorem}[The Decomposition Theorem~\cite{jeh2003scaling}] \label{thm:decomposition}
    \begin{align}
        \vpi_v&=\frac{1-\alpha}{\dout(v)}\sum_{u\in\Nout(v)}\vpi_u+\alpha\ve_v, \label{eqn:decom_SSPPR} \\
        \vpiinv_v&=(1-\alpha)\sum_{u\in\Nin(v)}\frac{\vpiinv_u}{\dout(u)}+\alpha\ve_v. \label{eqn:decom_STPPR}
    \end{align}
\end{theorem}

\header{PPR on Undirected Graphs.}
There are some additional properties for PPR on undirected graphs.
At a high level, as the main subject of \textit{spectral graph theory}, undirected graphs have many special algebraic properties.
As simple examples, we have $\bPi\D\vone=\P\D\vone=\D\vone=\A\vone$.
Also, random walks on undirected graphs are different from those on general directed graphs as they can be traversed oppositely.
Besides, it is well-known that a typical random walk (without restart) on a connected and non-bipartite undirected graph converges to the stationary distribution, in which each entry is proportional to the degree of the corresponding node.
As for PPR, the following key theorem forms the foundation for many PPR algorithms tailored to undirected graphs:
\begin{theorem}[Symmetry of PPR on Undirected Graphs~\protect{\cite[Lemma~1]{avrachenkov2013choice}}] \label{thm:symmetry}
$d(s)\cdot\vpi_s(t)=d(t)\cdot\vpi_t(s)$ holds for all $s,t\in V$.
\end{theorem}

Of course, these properties of PPR are far from comprehensive.
Interested readers are referred to \cite{bianchini2005inside,langville2004deeper} for detailed explanations of the mathematical properties of PPR.
\Tbl~\ref{tab:notations} lists the frequently used notations in this paper.

\section{Basic Techniques} \label{sec:basic_techniques}

In this section, we elaborate on five basic techniques for computing PPR: the \textit{Monte Carlo method}~\cite{fogaras2005towards}, \textit{Power Iteration}~\cite{brin1998anatomy}, \textit{Forward Push}~\cite{andersen2006local,andersen2007pagerank}, \textit{Reverse Power Iteration}~\cite{andersen2007local,andersen2008local}, and \textit{Backward Push}~\cite{andersen2007local,andersen2008local}.
They are the building blocks of more advanced algorithms for PPR computation.
We give detailed descriptions and analyses of these techniques, as well as a comparison of their pros and cons.

\subsection{The Monte Carlo Method} \label{sec:basic_algo_MC}

The Monte Carlo method (MC) is a classic approach for calculating sample-based estimates of desired values.
Inspired by the interpretation of PPR based on $\alpha$-discounted random walks, we can readily approximate the PPR vector $\vpi_s$ using MC.
Albeit seemingly na{\"{i}}ve, MC can obtain unbiased and accurate PPR estimates efficiently with high probability, and it can potentially break the theoretical lower bounds for deterministic algorithms.
Another latent advantage of MC is its inherent suitability for parallelization.

The fundamental idea of MC is to simulate $\alpha$-discounted random walks from $s$ and use the empirical termination distribution as the approximation for $\vpi_s$.
This provides a natural way to answer the SSPPR query.
Suppose that we generate $W$ random walks independently, which takes expected $\Theta(W/\alpha)$ time.
By \textit{Chernoff bound}, setting $W=\Theta\left(\log(n/\delta)\big/\eps^2\right)$ suffices to answer the probabilistic SSPPR query with absolute error bounds and setting $W=\Theta\left(\log(n/\delta)\big/\left(\eps^2\mu\right)\right)$ suffices for that with relative error bounds~\cite{fogaras2005towards}.

\header{Variants of MC.}
To improve the efficiency, numerous variants of MC have been proposed, with different approaches to generate random walks and extract information from the resultant walks.
Avrachenkov et al.~\cite{avrachenkov2007monte} suggest several versions of MC, in particular one that takes into account the entire traversed path of the random walks, called \textit{``MC complete path.''}
Wei et al.~\cite{wei2018topppr} propose a novel type of random walk called \textit{$\left(\sqrt{1-\alpha}\right)$-walk}, which has lower variances.
Liao et al.~\cite{liao2022efficient} present an algorithm for sampling a \textit{random spanning forest} of the graph.
They prove that such a sampling takes considerably less time than sampling $n$ random walks, especially for small $\alpha$.
They also argue that for SSPPR and STPPR computation, such a forest sampling provides information comparable to that of $n$ random walk samplings.

\subsection{Power Iteration}

Power Iteration (PI) is a basic iterative approach for computing the whole PPR vector, as initially introduced in the seminal paper by Google~\cite{brin1998anatomy}.
In fact, it is a direct adaptation of the \textit{Jacobi Method} for solving linear equations in numerical linear algebra (see, e.g., \cite[Section 4.1]{saad2003iterative}), where $\vpi_s$ is viewed as the solution to $\vpi_s=(1-\alpha)\P\vpi_s+\alpha\ve_s$ (\Eqn~\eqref{eqn:PPR_def}).

PI uses the following iteration formula to compute estimates $\vepi_s^{(0)},\vepi_s^{(1)},\ldots,\vepi_s^{(T)}$ for $\vpi_s$, where $T$ is a parameter:
\begin{align}
    \vepi_s^{(0)}=\ve_s,\quad \,\, \vepi_s^{(L+1)}=(1-\alpha)\P\vepi_s^{(L)}+\alpha\ve_s, \,\,\forall L\ge0. \label{eqn:SPI_iteration}
\end{align}
Here, $\vepi_s^{(L)}$ equals the probability distribution of an RWR starting from $s$ after $L$ steps.
Note that since $\P$ has exactly $m$ nonzero entries, \Eqn~\eqref{eqn:SPI_iteration} can be computed in $O(m)$ time.
Furthermore, it can be shown that one can set $T=\Theta\left(\frac{1}{\alpha}\log\frac{1}{\eps}\right)$ to ensure that $\left\lVert\vepi_s^{(T)}-\vpi_s\right\rVert_1\le\eps$.
Thus, by returning $\vepi_s^{(T)}$ as an estimate for $\vpi_s$, PI solves the SSPPR query with $\ell_1$-error bound $\eps$ in $O(mT)=O\left(\frac{m}{\alpha}\log\frac{1}{\eps}\right)$ time.

\header{Cumulative Power Iteration.}
We additionally introduce a variant of PI since it can be seen as the prototype of Forward Push (see \Sec~\ref{sec:forward_push}).
We call it \textit{Cumulative Power Iteration (CPI)}~\cite{yoon2018tpa} to distinguish it from \textit{Standard Power Iteration (SPI)} discussed above.
CPI approaches $\vpi_s$ using \Eqn~\eqref{eqn:PPR_series}: $\vpi_s=\alpha\sum_{\ell=0}^{\infty}(1-\alpha)^{\ell}\P^{\ell}\ve_s$. To this end, we define:
\begin{align}
    \vr^{(L)}&=(1-\alpha)^L\P^L\ve_s, \label{eqn:def_residue_vector} \\
    \vepi_s^{(L)}&=\alpha\sum_{\ell=0}^{L-1}\vr^{(\ell)}=\alpha\sum_{\ell=0}^{L-1}(1-\alpha)^{\ell}\P^{\ell}\ve_s. \label{eqn:def_reserve_vector}
\end{align}
We can compute $\vr^{(L)}$ and $\vepi_s^{(L)}$ iteratively as follows:
\begin{alignat}{2}
    \vr^{(0)}&=\ve_s,&\quad\vr^{(L+1)}&=(1-\alpha)\P\vr^{(L)},  \label{eqn:CPI_iteration_1} \\
    \vepi_s^{(0)}&=\vzero,&\quad\vepi_s^{(L+1)}&=\vepi_s^{(L)}+\alpha\vr^{(L)}. \label{eqn:CPI_iteration_2}
\end{alignat}
Similar to SPI, each iteration of CPI also takes $O(m)$ work.

Again, we choose a parameter $T$, compute $\vepi_s^{(0)}$ through $\vepi_s^{(T)}$ and return $\vepi_s^{(T)}$ as the final estimate.
The $\ell_1$-error between $\vepi_s^{(T)}$ and $\vpi_s$ can be precisely calculated as
\begin{align*}
    &\left\lVert\vepi_s^{(T)}-\vpi_s\right\rVert_1=\left\lVert\alpha\sum_{\ell=T}^{\infty}(1-\alpha)^{\ell}\P^{\ell}\ve_s\right\rVert_1 \\
    =&\alpha\sum_{\ell=T}^{\infty}(1-\alpha)^{\ell}\left\lVert\P^{\ell}\ve_s\right\rVert_1=\alpha\sum_{\ell=T}^{\infty}(1-\alpha)^{\ell}=(1-\alpha)^{T}.
\end{align*}
Thus, we have $\left\lVert\vepi_s^{(T)}-\vpi_s\right\rVert_1\le\eps$ if $T=\Theta\left(\frac{1}{\alpha}\log\frac{1}{\eps}\right)$, so CPI guarantees an $\ell_1$-error of $\eps$ in $O\left(\frac{m}{\alpha}\log\frac{1}{\eps}\right)$ time.

\header{Reserve and Residue Vectors in CPI.}
By \Eqns~\eqref{eqn:def_residue_vector} and \eqref{eqn:def_reserve_vector}, the following \textit{invariant} holds for any $L\ge0$:
\begin{align}
    \vpi_s=\vepi_s^{(L)}+\bPi\vr^{(L)}. \label{eqn:CPI_invariant}
\end{align}
Motivated by this, $\vr^{(L)}$ can be regarded as the \textit{residue vector} associated with $\vepi_s^{(L)}$, the \textit{reserve vector}.
The reserve vector $\vepi_s^{(L)}$ is always an underestimate of $\vpi_s$, where the associated residue vector determines the error between them.

\header{Push-Based Interpretation of CPI.}
Let us inspect \Eqns~\eqref{eqn:CPI_iteration_1} and \eqref{eqn:CPI_iteration_2} more carefully.
By examining them from the perspective of an individual node $v\in V$, we find that, in effect, $\alpha$ fraction of $\vr^{(L)}(v)$ (its \textit{residue}) is converted to $\vepi_s^{(L+1)}(v)$ (its \textit{reserve}) and the rest $(1-\alpha)$ fraction is propagated evenly to the out-neighbors of $v$ in $\vr^{(L+1)}$ (their residues) in each iteration.
This process can be defined as a \textit{push} operation, an important atomic operation widely used in recent algorithms.
From this point of view, CPI can be seen as performing pushes \textit{synchronously} to all nodes in $V$ in each iteration.

In practice, the two versions of PI have nearly identical effects, in the sense that they both solve the SSPPR query with $\ell_1$-error bound $\eps$ in time $O\left(\frac{m}{\alpha}\log\frac{1}{\eps}\right)$.
A minor technical distinction is that all estimated vectors in SPI are valid probability distributions, whereas, in CPI, they are underestimates of $\vpi_s$.
Besides, the implementation of SPI is simpler as it does not need to maintain the additional residue vector.
Thus, SPI may be more efficient in practice.
Nevertheless, these two algorithms are often not finely distinguished and are collectively referred to as PI in the literature.

\subsection{Forward Push} \label{sec:forward_push}

Forward Push (FP)~\cite{andersen2006local,andersen2007pagerank} is a \textit{local-push} method that can answer SSPPR queries without having to scan the whole graph.
In essence, FP can be viewed as an \textit{iterative coordinate solver}~\cite{fountoulakis2019variational} and also a variant of the \textit{Gauss-Seidel} method~\cite{chen2023accelerating}.
For our purpose, it is best understood as a local version of CPI, as explained thoroughly in \cite{wu2021unifying} and briefly below.

Recall that CPI performs pushes to all nodes synchronously in each iteration.
On the other hand, FP modifies this procedure and performs pushes \textit{asynchronously}.
It only performs one push operation for a single node at a time and allows the nodes to be pushed in a less structured order.
Intuitively, nodes with larger reserve values should take priority to be pushed.

\begin{figure*}[!t]
\centering
\subfloat[Initial State]{\includegraphics[width=0.2\textwidth]{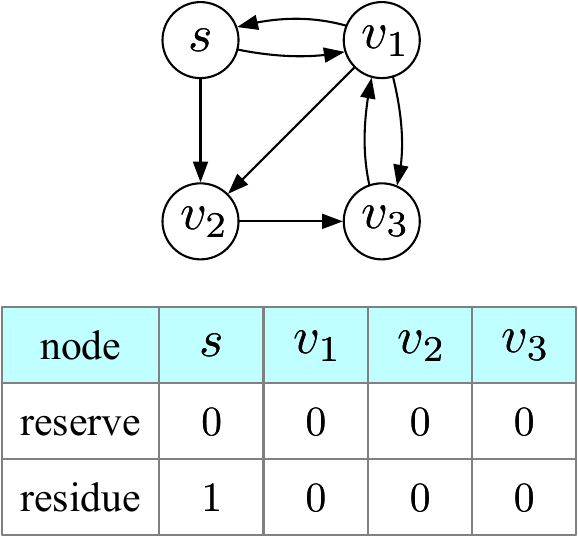}%
}
\hfil
\subfloat[Step 1]{\includegraphics[width=0.2\textwidth]{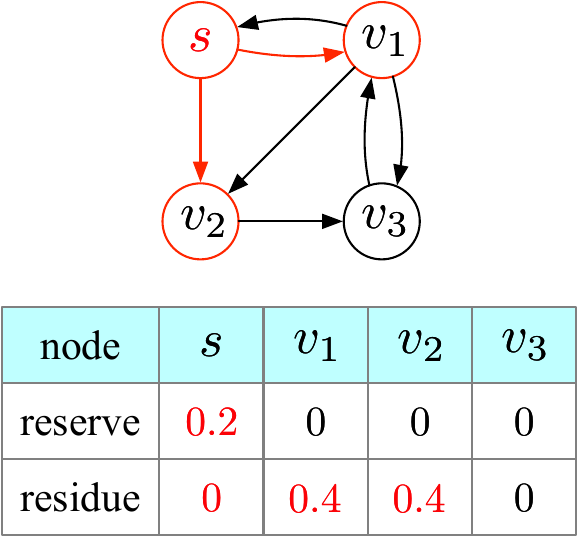}%
}
\hfil
\subfloat[Step 2]{\includegraphics[width=0.2\textwidth]{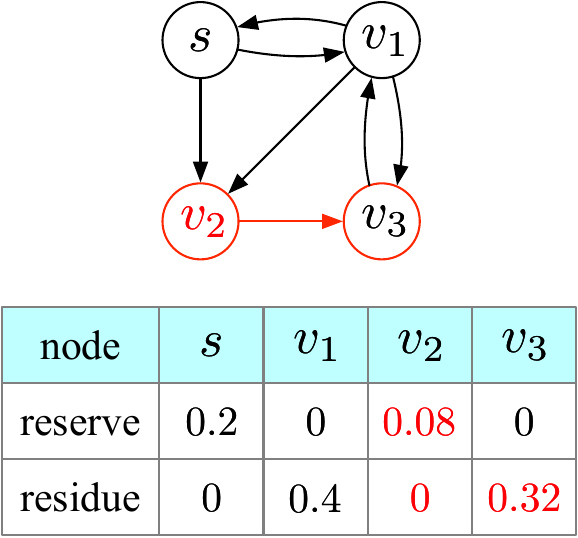}%
}
\hfil
\subfloat[Step 3]{\includegraphics[width=0.2\textwidth]{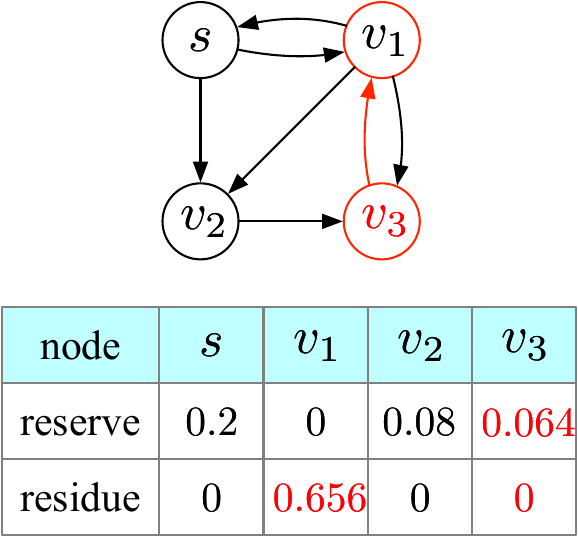}%
}
\caption{A running example of Forward Push on a toy graph.
        $s$ is the source node, $\alpha$ is set to $0.2$ and $\rmaxf$ is set to $0.3$.
        Each step stands for a single push operation and updated information is marked in red.}
\label{fig:running_example_FP}
\end{figure*}

Specifically, the pseudocode of FP is shown in \Alg~\ref{alg:FP}, which receives an extra parameter $\rmaxf$ as input.
FP initializes the reserve and residue vectors as $\vepif_s=\vzero,\vrf_s=\ve_s$ in Line~\ref{line:FP_init}.
After that, FP performs push operations to nodes $v$ with $\frac{\vrf_s(v)}{\dout(v)}>\rmaxf$ in an arbitrary order in Lines~\ref{line:FP_loop_begin} to~\ref{line:FP_loop_end}.
When no nodes $v$ satisfy the condition $\frac{\vrf_s(v)}{\dout(v)}>\rmaxf$ in Line~\ref{line:FP_loop_begin}, FP terminates the loop and returns the reserve and residue vectors in Line~\ref{line:FP_return}.
\Fig~\ref{fig:running_example_FP} illustrates a running example of FP.
In the following, we analyze the properties of this simple algorithm.

\begin{algorithm}[t]
    \DontPrintSemicolon
    \caption{Forward Push~\cite{andersen2006local,andersen2007pagerank}}
    \label{alg:FP}
    \KwIn{graph $G=(V,E)$, source node $s$, teleportation probability $\alpha$, threshold $\rmaxf$}
    \KwOut{reserve vector $\vepif_s$ and residue vector $\vrf_s$}
    $\vepif_s\gets\vzero,\vrf_s\gets\ve_s$ \; \label{line:FP_init}
    \While{$\exists v\in V$ {\rm s.t.} $\frac{\vrf_s(v)}{\dout(v)}>\rmaxf$ \label{line:FP_loop_begin}}
    {
        pick an arbitrary node $v$ with $\frac{\vrf_s(v)}{\dout(v)}>\rmaxf$ \;
        $r\gets\vrf_s(v),\vrf_s(v)\gets0$ \;
        $\vepif_s(v)\gets\vepif_s(v)+\alpha r$ \;
        \For{{\rm each} $u\in\Nout(v)$}
        {
            $\vrf_s(u)\gets\vrf_s(u)+\frac{1-\alpha}{\dout(v)}\cdot r$ \; \label{line:FP_loop_end}
        }
    }
    \Return $\vepif_s$ and $\vrf_s$ \; \label{line:FP_return}
\end{algorithm}

\header{Invariant.}
Recall that CPI satisfies the invariant $\vpi_s=\vepi_s^{(L)}+\bPi\vr^{(L)}$ (\Eqn~\eqref{eqn:CPI_invariant}).
Likewise, the following invariant holds before and after each push operation in FP:
\begin{align}
    \vpi_s=\vepif_s+\bPi\vrf_s. \label{eqn:FP_invariant}
\end{align}
This key invariant can be justified using induction and the decomposition theorem (\Eqn~\eqref{eqn:decom_SSPPR}).

\header{Complexity.}
The total cost of performing pushes dominates the complexity of FP.
Note that Line~\ref{line:FP_loop_begin} constrains that each pushed node $v$ must satisfy $\frac{\vrf_s(v)}{\dout(v)}>\rmaxf$, so one push operation for $v$ increases $\left\lVert\vepif_s\right\rVert_1$ by at least $\alpha\cdot\dout(v)\cdot\rmaxf$ at the cost of $O\big(\dout(v)\big)$ time.
Therefore, $\alpha\cdot\rmaxf$ can be viewed as the ``unit cost-benefit'' of the push operations in terms of increasing $\left\lVert\vepif_s\right\rVert_1$.
On the other hand, as the invariant (\Eqn~\eqref{eqn:FP_invariant}) implies, $\vepif_s$ is always an underestimate of $\vpi_s$, and thus $\left\lVert\vepif_s\right\rVert_1\le\left\lVert\vpi_s\right\rVert_1=1$.
Hence, the complexity of pushes is bounded by this upper bound of $\left\lVert\vepif_s\right\rVert_1$ divided by the unit cost-benefit $\alpha\cdot\rmaxf$, which equals $O\left(1\Big/\left(\alpha\cdot\rmaxf\right)\right)$.

\header{Error Bound.}
The condition in Line~\ref{line:FP_loop_begin} also implies that, after termination, the residue vector satisfies $\vrf_s(v)\le\rmaxf\cdot\dout(v),\forall v\in V$.
This implies that $\left\lVert\vepif_s-\vpi_s\right\rVert_1=\left\lVert\bPi\vrf_s\right\rVert_1\le\left\lVert\bPi\cdot\rmaxf\cdot\D\vone\right\rVert_1=m\cdot\rmaxf$.
In light of this, one can set $\rmaxf=\eps/m$ and invoke FP to answer the SSPPR query with $\ell_1$-error bound $\eps$. Under such settings, the time complexity becomes $O\big(m/(\alpha\eps)\big)$.

\subsection{Reverse Power Iteration}

Regarding STPPR, it is natural to ask whether a method called reverse power iteration (RPI) exists that works analogously to PI.
Fortunately, the answer is positive.
Similar to PI, RPI makes use of \Eqn~\eqref{eqn:STPPR_series}: $\vpiinv_t=\alpha\sum_{\ell=0}^{\infty}(1-\alpha)^{\ell}\left(\P^{\top}\right)^{\ell}\ve_t$.
Following CPI and using iterations akin to \Eqns~\eqref{eqn:CPI_iteration_1} and \eqref{eqn:CPI_iteration_2}, the estimate $\vepi_t^{(-T)}=\alpha\sum_{\ell=0}^{T-1}(1-\alpha)^{\ell}\left(\P^{\top}\right)^{\ell}\ve_t$ can be computed in $O(mT)$ time.
Here, $\vepi_t^{(-T)}$ denotes the $T$-th estimate for $\vpiinv_t$.

However, since $\P^{\top}$ is no longer column-stochastic as $\P$ does, the error between $\vepi_t^{(-T)}$ and $\vpiinv_t$ is a bit different.
It turns out that considering the absolute error (instead of $\ell_1$-error) between them is easier:
\begin{align*}
    &\left\lVert\vepi_t^{(-T)}-\vpiinv_t\right\rVert_{\infty}\le\left\lVert\alpha\sum_{\ell=T}^{\infty}(1-\alpha)^{\ell}\left(\P^{\top}\right)^{\ell}\vone\right\lVert_{\infty} \\
    =&\left\lVert\alpha\sum_{\ell=T}^{\infty}(1-\alpha)^{\ell}\vone\right\rVert_{\infty}=(1-\alpha)^T.
\end{align*}
To guarantee that $\left\lVert\vepi_t^{(-T)}-\vpiinv_t\right\rVert_{\infty}\le\eps$, it suffices to set $T=\left\lceil\log_{1-\alpha}\eps\right\rceil=\Theta\left(\frac{1}{\alpha}\log\frac{1}{\eps}\right)$.
In a word, RPI solves the STPPR query with absolute error bound $\eps$ deterministically in $O(mT)=O\left(\frac{m}{\alpha}\log\frac{1}{\eps}\right)$ time.

Analogous to CPI, RPI can also be viewed as performing \textit{reverse pushes} to all nodes synchronously in each iteration.
A reverse push operation to node $v$ also converts $\alpha$ fraction of its residue to its reserve but propagates the rest in a different way, which can be thought of as pushing according to the matrix $\P^{\top}$ instead of $\P$.
To explain, as the $(i,j)$-th entry of $\P^{\top}$ equals $\mathds{1}\big\{(v_j,v_i)\in E\big\}\big/\dout(v_j)$, the reverse push to $v$ should increment the residue of each $u\in\Nin(v)$ by $(1-\alpha)/\dout(u)\cdot\vrb(v)$.
RPI can also be implemented similarly to SPI, with the same asymptotic complexity.

\subsection{Backward Push}

Backward Push (BP)~\cite{andersen2007local,andersen2008local} is a local-push algorithm for answering the STPPR query.
It can be viewed as the reverse version of FP or the local version of RPI.
The framework of BP is extremely similar to that of FP, as shown in Alg.~\ref{alg:BP}.
It also receives a parameter $\rmaxb$ to determine which nodes to push, but the condition in Line~\ref{line:BP_loop} is slightly different.

\begin{algorithm}[t]
    \DontPrintSemicolon
    \caption{Backward Push~\cite{andersen2007local,andersen2008local}}
    \label{alg:BP}
    \KwIn{graph $G=(V,E)$, target node $t$, teleportation probability $\alpha$, threshold $\rmaxb$}
    \KwOut{reserve vector $\vepib_t$ and residue vector $\vrb_t$}
    $\vepib_t\gets\vzero,\vrb_t\gets\ve_t$ \;
    \While{$\exists v\in V$ {\rm s.t.} $\vrb_t(v)>\rmaxb$ \label{line:BP_loop}}
    {
        pick an arbitrary node $v$ with $\vrb_t(v)>\rmaxb$ \;
        $r\gets\vrb_t(v),\vrb_t(v)\gets0$ \;
        $\vepib_t(v)\gets\vepib_t(v)+\alpha r$ \;
        \For{{\rm each} $u\in\Nin(v)$}
        {
            $\vrb_t(u)\gets\vrb_t(u)+\frac{1-\alpha}{\dout(u)}\cdot r$ \;
        }
    }
    \Return $\vepib_t$ and $\vrb_t$ \;
\end{algorithm}

\header{Invariant.}
Analogous to FP, BP also admits an invariant between the \textit{reserve vector} $\vepib_t$ and the \textit{residue vector} $\vrb_t$:
\begin{align}
    \vpiinv_t=\vepib_t+\bPi^{\top}\vrb_t. \label{BP_invariant}
\end{align}
The proof of this invariant also relies on induction and the decomposition theorem (\Eqn~\eqref{eqn:decom_STPPR}).

\header{Complexity.}
Similarly, the complexity of BP is also dominated by performing pushes.
Following the reasoning in FP, we find that a push operation on node $v$ increases $\vepib_t(v)$ by at least $\alpha\cdot\rmaxb$ in $O\big(\din(v)\big)$ time.
By considering each node $v\in V$ separately, we attain the total complexity of
\begin{align}
    O\left(\sum_{v\in V}\frac{\vpiinv_t(v)}{\alpha\cdot\rmaxb}\cdot\din(v)\right)=O\left(\frac{\sum_{v\in V}\vpiinv_t(v)\cdot\din(v)}{\alpha\cdot\rmaxb}\right). \label{eqn:BP_complexity}
\end{align}
To derive a simpler complexity, we can consider BP's \textit{average complexity} over all target nodes $t\in V$, which equals
\begin{align*}
    O\left(\frac{1/n}{\alpha\cdot\rmaxb}\sum_{v\in V}\din(v)\sum_{t\in V}\vpi_v(t)\right)=O\left(\frac{m/n}{\alpha\cdot\rmaxb}\right).
\end{align*}

\header{Error Bound.}
As Line~\ref{line:BP_loop} implies, after termination, the residue vector satisfies $\vrb_t(v)\le\rmaxb,\,\,\forall v\in V$.
It follows that
\begin{align*}
    &\left\lVert\vepib_t-\vpiinv_t\right\rVert_{\infty}=\left\lVert\bPi^{\top}\vrb_t\right\rVert_{\infty} \\
    \le&\left\lVert\bPi^{\top}\cdot\rmaxb\cdot\vone\right\rVert_{\infty}=\rmaxb\cdot\lVert\vone\rVert_{\infty}=\rmaxb.
\end{align*}
Therefore, one can invoke BP to answer the STPPR query with absolute error bound $\eps$ by setting $\rmaxb=\eps$, with an average complexity of $O\left(m/n/(\alpha\eps)\right)$.

\subsection{FP and BP on Undirected Graphs}

FP and BP exhibit some additional properties on undirected graphs.
In particular, using the specific properties discussed at the end of \Sec~\ref{sec:properties}, we can simplify the error bound of FP and the complexity bound of BP as follows.

\header{FP on Undirected Graphs.}
The error bound of FP can be specific to each node on undirected graphs.
To see this, note that $\vepif_s=\vpi_s-\bPi\vrf_s$ is lower bounded by
\begin{align*}
    \vpi_s-\bPi\cdot\rmaxf\cdot\D\vone=\vpi_s-\rmaxf\cdot\D\vone,
\end{align*}
where we used $\bPi\D\vone=\D\vone$ for undirected graphs.
Thus,
\begin{align*}
    &\vpi_s(v)-\rmaxf\cdot d(v)\le\vepif_s(v)\le\vpi_s(v),\quad\forall v\in V, \\ \Leftrightarrow&\frac{\vpi_s(v)}{d(v)}-\rmaxf\le\frac{\vepif_s(v)}{d(v)}\le\frac{\vpi_s(v)}{d(v)},\quad\forall v\in V.
\end{align*}
This type of result is called \textit{degree-normalized absolute error}, which means that the values $\vpi_s(v)/d(v)$ are estimated with absolute error bounds.
We conclude that FP can solve the SSPPR query with degree-normalized absolute error bound $\eps$ in $O\big(1/(\alpha\eps)\big)$ time on undirected graphs by setting $\rmaxf=\eps$.
Although this type of error measure allows larger errors for nodes with larger degrees, it is useful for some applications (e.g., for local graph clustering~\cite{andersen2006local}).

\header{BP on Undirected Graphs.}
\Eqn~\eqref{eqn:BP_complexity} can be directly simplified with the aid of \Thm~\ref{thm:symmetry} on undirected graphs, yielding $O\left(\sum_{v\in V}d(t)\cdot\vpi_t(v)\Big/\left(\alpha\cdot\rmaxb\right)\right)=O\big(d(t)/(\alpha\eps)\big)$.
Compared to $O\left(m/n/(\alpha\eps)\right)$ on directed graphs, this complexity is parameterized instead of average.

\subsection{Summary and Comparison of the Basic Techniques}

We summarize the considered query types and complexities of the five basic techniques for different error bounds in \Tbl~\ref{tab:basic_algo}.
We remark that MC applies to probabilistic queries, while others pertain to deterministic ones.
Also, the bounds of BP are averaged over all target nodes $t\in V$.
Furthermore, we highlight the complexities derived above in red.
The remaining complexities in black are derived using conversions between the error measures (discussed at the end of \Sec~\ref{sec:def_PPR_queries}) and thus may be pessimistic.

\begin{table*}[!t]
\renewcommand{\arraystretch}{1.5}
\caption{Summary of Basic Techniques (\textcolor{black}{Black} Complexities Are Derived using Error-Bound Conversions and May Be Inferior)} \label{tab:basic_algo}
\centering
    \begin{tabular}{ScScScScScSc}
        \toprule
        \multirow{2}{*}{\textbf{Technique}} & \multirow{2}{*}{\textbf{Query Type}} & \multicolumn{4}{c}{\textbf{Error Measures}} \\
         & & $\boldsymbol{\ell_1}$ & {\textbf{absolute}} & {\textbf{relative}} & {\textbf{degree-normalized absolute}} \\
        \midrule
    	MC~\cite{fogaras2005towards} & SSPPR (probabilistic) & $O\left(\dfrac{n^2\log(n/\delta)}{\alpha\eps^2}\right)$ & \textcolor{red}{$O\left(\dfrac{\log(n/\delta)}{\alpha\eps^2}\right)$} & \textcolor{red}{$O\left(\dfrac{\log(n/\delta)}{\alpha\eps^2\mu}\right)$} & $O\left(\dfrac{\log(n/\delta)}{\alpha\eps^2}\right)$ \\
    	PI~\cite{brin1998anatomy} & SSPPR & \textcolor{red}{$O\left(\dfrac{m}{\alpha}\log\dfrac{1}{\eps}\right)$} & $O\left(\dfrac{m}{\alpha}\log\dfrac{1}{\eps}\right)$ & $O\left(\dfrac{m}{\alpha}\log\dfrac{1}{\eps\mu}\right)$ & $O\left(\dfrac{m}{\alpha}\log\dfrac{1}{\eps}\right)$ \\
    	FP~\cite{andersen2007pagerank} & SSPPR & \textcolor{red}{$O\left(\dfrac{m}{\alpha\eps}\right)$} & $O\left(\dfrac{m}{\alpha\eps}\right)$ & $O\left(\dfrac{m}{\alpha\eps\mu}\right)$ & \textcolor{red}{$O\left(\dfrac{1}{\alpha\eps}\right)$} \\
    	RPI~\cite{andersen2008local} & STPPR & $O\left(\dfrac{m}{\alpha}\log\dfrac{n}{\eps}\right)$ & \textcolor{red}{$O\left(\dfrac{m}{\alpha}\log\dfrac{1}{\eps}\right)$} & $O\left(\dfrac{m}{\alpha}\log\dfrac{1}{\eps\mu}\right)$ & $O\left(\dfrac{m}{\alpha}\log\dfrac{1}{\eps}\right)$ \\
    	BP~\cite{andersen2008local} & STPPR & $O\left(\dfrac{m}{\alpha\eps}\right)$ (average) & \textcolor{red}{$O\left(\dfrac{m/n}{\alpha\eps}\right)$} (average) & $O\left(\dfrac{m/n}{\alpha\eps\mu}\right)$ (average) & $O\left(\dfrac{m/n}{\alpha\eps}\right)$ (average) \\
    	\bottomrule
    \end{tabular}
\end{table*}

Generally speaking, PI and RPI are inefficient for massive graphs, as they need to scan the whole graph in each iteration.
However, their dependence on $\eps$, $O\big(\log(1/\eps)\big)$, is the mildest among these methods.
On the contrary, MC does not need to scan the graph, but its dependence on $\eps$ is $O\left(1\big/\eps^2\right)$, a considerable factor when $\eps$ is small.
The local-push methods seem to be a compromise between these two extremes: FP and BP work by locally exploring the graph, and their dependence on $\eps$ is $O(1/\eps)$.
MC and the local-push methods are the key ingredients for designing sublinear algorithms.

Let us take a closer look at FP and BP.
Despite their highly similar approaches, their properties have some key differences:
\begin{itemize}[leftmargin=*]
    \item In terms of error bounds, BP can elegantly provide absolute error guarantees, while FP only provides degree-normalized absolute error guarantees on undirected graphs.
    Note that although FP can provide $\ell_1$-error guarantee on directed graphs, in this case, its complexity becomes proportional to $m$, and thus, it becomes nonlocal and inferior to PI.
    \item In terms of complexity, FP runs in $O\left(1\Big/\left(\alpha\cdot\rmaxf\right)\right)$ time, while BP's complexity on directed graphs cannot be expressed compactly.
    BP exhibits a simple form of complexity only when considering average complexity or when applied to undirected graphs.
    In these two special cases, its complexity becomes $O\left(m/n\Big/\left(\alpha\cdot\rmaxb\right)\right)$ and $O\left(d(t)\Big/\left(\alpha\cdot\rmaxb\right)\right)$, resp.
\end{itemize}

\section{Overview of PPR Algorithms} \label{sec:overview}

The remainder of this paper focuses on more recent and advanced PPR algorithms.
Although these algorithms cover various query types, error measures, and computing settings, they can be roughly divided into two categories by their methodologies: PPR algorithms based on \textit{PI and linear algebraic techniques} and on \textit{local-push methods and MC}.

\begin{enumerate}[leftmargin=*]
    \item PPR algorithms based on PI and linear algebraic techniques interpret PPR as the solution to \Eqn~\eqref{eqn:PPR_def}.
    They harness various \textit{numerical methods} and \textit{optimization methods} to expedite the solving process.
    Some works also leverage the structures of real-world graphs to simplify the problem.
    In general, these algorithms can obtain exact PPR values or approximate ones with tiny errors, but they need at least linear time in the graph size.
    Thus, they are more suitable for precomputing high-precision PageRank vectors or PPR matrices for small graphs.
    These works mainly consider improving the convergence speed and utilizing parallel/distributed settings to achieve higher scalability.
    \item PPR algorithms based on local-push methods and MC rely on the random-walk-based interpretations of PPR.
    They use a combination of local-push methods and MC to derive approximate PPR values in a local manner.
    Although they cannot obtain extremely accurate results, some of them achieve sublinear complexities, which are more suitable for online queries and are particularly desirable for certain queries.
    These works mainly consider combining and optimizing the techniques of FP, BP, and MC to reduce the complexity, and some works also design effective precomputation to diminish the query cost.
\end{enumerate}

We summarize the PPR algorithms surveyed below in \Tbls~\ref{tab:summary} and \ref{tab:summary_special} at the end of this paper.

\section{Single-Source PPR Algorithms} \label{sec:SSPPR_algorithms}

The majority of recent PPR research concentrates on the SSPPR queries.
We introduce them below based on the taxonomy of their techniques.
For ease of presentation, unless explicitly stated otherwise, we assume that $\alpha$ is constant in the rest of this paper.
We use $\tO$ to denote the big-$O$ notation ignoring polylogarithmic factors.

\subsection{SSPPR Algorithms Based on PI and Algebraic Techniques} \label{sec:adv_alg_SSPPR_PI}

This type of SSPPR algorithm starts from the algebraic definition of PPR (\Eqn~\eqref{eqn:PPR_def}) and applies several advanced linear algebraic or optimization techniques to compute PPR vectors with small errors efficiently.
However, their complexities are generally at least linear in the size of the graph.

\header{PageRank Vector Computation.}
Several earlier works~\cite{kamvar2003extrapolation,broder2004efficient,kamvar2004adaptive} try to improve the efficiency of PI for computing the PageRank vector by leveraging linear algebraic techniques and the structural properties of real-world graphs.
We refer to a detailed survey by Berkhin~\cite{berkhin2005survey} and the references therein for these algorithms.
Additionally, Avrachenkov et al.~\cite{avrachenkov2007monte} use MC and its variants to estimate the PageRank vector, and Wu and Wei~\cite{wu2010arnoldi} apply two advanced linear algebraic methods, \textbf{Arnoldi} and \textbf{GMRES}, to compute the PageRank vector.

Some earlier works for SSPPR include \textbf{B\_LIN} and \textbf{NB\_LIN}~\cite{tong2006fast}, which leverage \textit{graph partitioning} and \textit{low-rank approximation}, and \textbf{RPPR} and \textbf{BRPPR}~\cite{gleich2007approximating}, which use SPI on restricted subgraphs and expand the subgraph iteratively.
Additionally, Jeh and Widom~\cite{jeh2003scaling} put forward a basic \textit{dynamic programming} algorithm for computing SSPPR based on the decomposition theorem (\Eqn~\eqref{eqn:decom_SSPPR}).
Sarl{\'{o}}s et al.~\cite{sarlos2006randomize} later improve this dynamic programming algorithm by applying \textit{Count-Min Sketch} to the intermediate vectors.
Under the framework of \textit{two-phase algorithms}, the algorithms need to construct a database in the first precomputation phase and can only access the database instead of the graph in the second query phase.
Their method reduces the database space from $O\big(n\log n\log(1/\delta)\big/\eps^2\big)$ in their previous work~\cite{fogaras2005towards} to $O\big(n\log(1/\delta)/\eps\big)$, where $\eps$ denotes the absolute error bound.
They demonstrate that this approach achieves optimal space usage for the database.

\textbf{FastPPV}, proposed by Zhu et al.~\cite{zhu2013incremental}, supports a dynamic trade-off between efficiency and accuracy at query time based on \textit{scheduled approximation}.
Its basic idea is to \textit{partition and prioritize} all paths from $s$ so that each iteration corresponds to a set of paths, and the critical paths can be considered in earlier iterations.
To this end, FastPPV first carefully selects a set of hub nodes $H$ and conceptually categorizes the paths according to the number of hub nodes on the paths, and then precomputes the PPR vector of each hub node in a small subgraph.
At query time, these vectors form the building blocks for the iterative computation, and the $\ell_1$-error bound decreases exponentially.

Maehara et al.~\cite{maehara2014computing} propose an SSPPR algorithm that exploits the graph structure and the iterative method GMRES.
It first performs \textit{core-tree decomposition} to transform the graph into a core and a small graph, then runs \textit{LU decomposition} to the small graph, and finally applies GMRES to the whole graph, using the previous result as a \textit{preconditioner}.

Jung et al.~\cite{shin2015bear,jung2016random} propose \textbf{BEAR}, a method that preprocesses several matrices and utilizes them via a \textit{block elimination approach}.
They propose two versions of BEAR, \textbf{\textsc{BEAR-Exact}} and \textbf{\textsc{BEAR-Approx}}, which give exact and approximate results, resp.
The overall idea is to reduce the overhead of computing $\big(\I-(1-\alpha)\P\big)^{-1}$ (cf. \Eqn~\eqref{eqn:PPR_inverse}) by using some linear algebraic techniques.
In the preprocessing phase, \textsc{BEAR-Exact} reorders the nodes so that the matrix $\I-(1-\alpha)\P$ after transformation contains several easy-to-invert sub-matrices and uses the \textit{Schur complement method} and \textit{LU decomposition} to prepare auxiliary matrices.
In the query phase, \textsc{BEAR-Exact} applies the \textit{block elimination method} to compute the PPR vector efficiently while exploiting the preprocessed matrices.
\textsc{BEAR-Approx} further heuristically omits some small entries in the preprocessed matrices to improve efficiency while sacrificing little accuracy in practice.

The follow-up work \textbf{\textsc{BePI}} by Jung et al.~\cite{jung2017bepi} further improves over BEAR~\cite{jung2016random}.
\textsc{BePI} incorporates the advantages of preprocessing methods and iterative methods, i.e., high querying efficiency and scalability to large graphs, resp.
Specifically, \textsc{BePI} combines the \textit{block elimination method} (as used by BEAR) and GMRES, and it further applies some advanced techniques like \textit{reordering} the nodes using the \textit{hub-and-spike structure} of real-world graphs, \textit{sparsifying} the \textit{Schur complement}, and \textit{preconditioning} a linear system.
Compared to BEAR, \textsc{BePI} performs preprocessing much more efficiently and requires significantly less space.
It is worth mentioning that \textsc{BePI} offers $\ell_2$-error guarantee.

Proposed by Co{\c{s}}kun et al.~\cite{coskun2016efficient}, \textbf{\textsc{Chopper}} modifies SPI by taking a linear combination of $\vepi_s^{(0)},\vepi_s^{(1)},\dots,\vepi_s^{(T)}$ in \Eqn~\eqref{eqn:SPI_iteration} to be an estimate of $\vpi_s$.
They show that by assigning the coefficients of the combinations using \textit{Chebyshev polynomials}, the obtained sequence of estimates can converge more rapidly, and the complexity of each iteration remains $O(m)$.
Experimentally, \textsc{Chopper} can indeed reduce the number of required iterations significantly.

The \textbf{TPA} algorithm by Yoon et al.~\cite{yoon2018tpa} tries to approximate the PPR vector through \Eqn~\eqref{eqn:PPR_series}, as CPI does.
It divides the summation into three parts, the \textit{family part}, the \textit{neighbor part}, and the \textit{stranger part}, by limiting the index in three intervals from zero to infinity.
The three parts are tackled differently, and in particular, the neighbor part is approximated by a factor times the family part, with the intuition that the \textit{block-wise structure} of the real-world graphs causes the two parts to have similar distributions.
TPA requires $O(n+m)$ memory space and takes $O\big(m\log(1/\eps)\big)$ time for both preprocessing and online query, where $\eps$ is the $\ell_1$-error bound.

Chen et al.~\cite{chen2023accelerating} consider accelerating SSPPR computation using advanced optimization techniques.
They propose \textbf{FwdPushSOR}, an improved version of FP based on \textit{Successive Over Relaxation}~\cite[Section 4.1]{saad2003iterative}.
On undirected graphs, they formulate SSPPR computation as a convex optimization problem (cf. \cite{fountoulakis2019variational}) and apply two \textit{momentum-based acceleration methods} to it.
To achieve a specified $\ell_1$-error bound, their algorithms run in $\tO\big(m\big/\sqrt{\alpha}\big)$ time, with $\alpha$ regarded as a parameter, improving over PI's $\tO(m/\alpha)$ bound.

\subsection{SSPPR Algorithms Based on Local Push and MC} \label{sec:adv_alg_SSPPR_MC}

These algorithms mainly optimizes the vanilla FP algorithm and/or combines local-push techniques with MC for answering the SSPPR queries.
By applying local pushes and randomness, they are potentially able to achieve sublinear complexities.

Pavel Berkhin~\cite{berkhin2006bookmark} put forward an approximate SSPPR algorithm called \textbf{BCA}, which is similar to FP but utilizes precomputed \textit{hub vectors} to accelerate the computation.
Conceptually, the pushes are performed as in standard FP, but the residues get blocked once they enter a hub node, where the hub nodes are selected beforehand.
After performing the pushes, the obtained reserve vector is combined with the hub vectors of the hub nodes in accordance with the amount of blocked residues to obtain the final estimate.
Chakrabarti further proposes \textbf{\textsc{HubRank}}~\cite{chakrabarti2007dynamic} to improve over Berkhin's BCA~\cite{berkhin2006bookmark}.
For preprocessing, \textsc{HubRank} carefully chooses some hub nodes according to the history query log statistics and runs MC to compute their PPR vectors as indices.
At query time, the indices are used along with running PI over a small subgraph to yield the results.

\header{FP with Exponential Convergence.}
Chung and Zhao~\cite{chung2010sharp} consider optimizing FP when $\rmaxf$ is small.
They propose to divide the process into $\Theta\left(\log\left(1\big/\rmaxf\right)\right)$ phases with exponentially decreasing thresholds of $1/2,1/4,1/8,\dots$ until it gets as small as $\rmaxf$.
The obtained algorithm is shown to have an $O\left(m\log\left(1\big/\rmaxf\right)\right)$ complexity, similar to that of PI.
Although this complexity is not sublinear, its dependence on $\rmaxf$ is $O\left(\log\left(1\big/\rmaxf\right)\right)$.
Besides, Wu et al.~\cite{wu2021unifying} revisits FP by revealing the connection between CPI and FP.
They show that a common implementation of FP called \textbf{FIFO-FP}, which utilizes a queue to maintain the nodes to be pushed, actually admits a complexity of $O\left(\min\left(1\big/\rmaxf,m\log\frac{1}{m\cdot\rmaxf}\right)\right)$.
This implies that FIFO-FP is no worse than PI for answering the SSPPR query with $\ell_1$-error bound.
The authors also propose \textbf{PowerPush}, which unifies the advantages of both the local approach (FP) and the global approach (PI) to achieve superior efficiency and accuracy in practice.
Specifically, PowerPush exploits FP's asynchronous pushes to reduce the number of pushes and mimics PI to perform pushes in a cache-friendly way when most nodes have been involved. 

\header{FORA and Its Follow-Ups.}
\textbf{FORA} by Wang et al.~\cite{wang2017fora,wang2019efficient} is a representative algorithm for the SSPPR query with relative error guarantees.
It combines FP and MC in an elegant and nontrivial way, and this ``FP + MC'' framework has been adopted by many subsequent works~\cite{luo2019efficient,luo2019baton,wang2019parallelizing,lin2020index,luo2020improved,hou2021massively,mo2021agenda,wu2021unifying,liao2022efficient,liao2023efficient,hou2023personalized}, some of which are tailored for distributed environments (see \Sec~\ref{sec:distributed}).
Roughly speaking, FORA conducts random walks deterministically through FP before doing so randomly through MC, dramatically reducing the number of samples needed in MC.
Technically, instead of ignoring the residues after FP, FORA replaces the $\bPi$ matrix in the invariant of FP (\Eqn~\eqref{eqn:FP_invariant}) by PPR estimates obtained from MC.
For the MC phase, FORA simulates several random walks from each node in $V$, where the number of walks depends on their residues.
FORA's overall complexity is shown to be $O\left(\min\left\{\frac{1}{\eps}\sqrt{\frac{m\log(1/\delta)}{\mu}},\frac{\log(1/\delta)}{\eps^2\mu}\right\}\right)$.
Besides, a variant of FORA called \textbf{FORA+} is proposed to achieve higher empirical performance, which avoids the cost of generating random walks on the fly by storing precomputed ones as \textit{indices}.

Based on FORA, \textbf{ResAcc} is proposed by Lin et al.~\cite{lin2020index} to achieve higher efficiency without using any indices.
Motivated by the \textit{looping phenomenon} discovered by BCA~\cite{berkhin2006bookmark}, ResAcc first only performs pushes to the nodes that are no more than $h$-hops away from $s$ (except $s$ itself after the first push) so that the residue can accumulate at $s$ and the $(h+1)$-th layer, and then updates the reserves and residues for these nodes directly in accordance with the residue accumulated at $s$.
After that, the residues at the $(h+1)$-th layer are pushed, and FP is proceeded as usual, and finally, random walk samplings are performed as FORA does.
Owing to the accumulated residue, ResAcc can perform fewer pushes and random walks in practice.

Wu et al.~\cite{wu2021unifying} propose \textbf{SpeedPPR} to improve FORA by replacing FP in FORA's first phase with their PowerPush~\cite{wu2021unifying}.
SpeedPPR runs in $O\left(m\log\frac{\log(1/\delta)}{\eps^2\mu m}\right)$ time for relative error guarantees.
Although this bound is not sublinear in $m$, SpeedPPR's empirical efficiency surpasses that of previous algorithms, owing to PowerPush's combination of FP and PI.
Similar to FORA+, SpeedPPR can also store precomputed random walks as indices to achieve even higher performance.

\header{Variational Formulation.}
Fountoulakis et al.~\cite{fountoulakis2019variational} revisit FP by deriving a variational formulation to explicitly express the problem solved by FP as an optimization problem.
They show that FP is indeed an iterative coordinate solver for \Eqn~\eqref{eqn:PPR_def}.
They further formulate an \textit{$\ell_1$-regularized PageRank} problem that decouples the sparsity of the solution from the specific algorithms, whose solution meets the requirements for local clustering in \cite{andersen2006local}.
The authors then apply the \textit{iterative shrinkage-thresholding algorithm} to solve this problem and prove that it also solves the problem locally.
Although they do not derive a directly better complexity bound, their formulation reveals a promising connection between local graph processing and numerical optimization.
Whether one can apply some other optimization techniques to speed up the computation by a factor of $1/\sqrt{\alpha}$ is left as an open problem~\cite{fountoulakis2022open}.
By using \textit{conjugate directions} and exploiting \textit{sparsity}, \cite{martinez2023accelerated} gives some new bounds for the $\ell_1$-regularized PageRank problem but does not fully resolve the open problem.

\header{Optimizing the MC Technique.}
Liao et al.~\cite{liao2022efficient} present a new implementation of MC by establishing a connection between PPR and the weights of \textit{random spanning forests} of the graph.
Armed with a new matrix forest theorem, they devise an approach for sampling a random spanning forest via simulating \textit{loop-erased $\alpha$-discounted random walks}.
Such a sampling can substitute $n$ random walk samplings under FORA's framework, and it takes expected $O\left(\sum_{v\in V}\pi(v,v)\right)$ time, which turns out to be significantly smaller than the corresponding $\Theta(n)$ complexity on real-world graphs.
By modifying FORA's FP phase and replacing the random walk samplings in the MC phase with their forest samplings, the authors propose \textbf{FORAL}, an efficient SSPPR algorithm that demonstrates robust performance for small $\alpha$.

A subsequent work by Liao et al.~\cite{liao2023efficient} introduces two innovative techniques to reduce the variances of the MC estimators.
The first technique performs several iterations of PI to MC estimates, leading to more accurate results.
The second one, called the \textit{progressive sampling} method, modifies the sampling scheme of MC by taking historical samplings into account.
The authors give theoretical analyses of these techniques' effects on variance reduction.
They combine these techniques with their previous \textit{spanning forest sampling} technique~\cite{liao2022efficient} to propose novel and practically efficient SSPPR algorithms.

\header{FP with Bounded Sensitivity.}
Epasto et al. \cite{epasto2022differentially} consider computing approximate PPR vectors with provably bounded \textit{sensitivity} to the edges in the undirected graph.
To this end, they modify FP and restrict the maximum amount of residues being pushed along each edge.
They show that this approach ensures low sensitivity and achieves a similar approximation quality as FP if the graph has a large minimum degree.

\section{Single-Target PPR Algorithms} \label{sec:STPPR_algorithms}

So far, relatively few studies exist for the STPPR queries, which mainly rely on reverse local pushes.
A basic difficulty is that MC cannot be used directly for STPPR computation.

Lofgren and Goel~\cite{lofgren2013personalized} propose an implementation of BP using a \textit{Fibonacci heap}.
They prove that the complexity of this implementation can be alternatively bounded by $O\big(m\log(1/\eps)\big)$, matching the complexity of RPI.

Wang et al.~\cite{wang2020personalized} propose \textbf{RBS}, which applies randomness to the original BP.
The key insight of RBS is that the cost of BP can be greatly reduced by only performing pushes to a randomly chosen subset of the in-neighbors of a node.
For the STPPR query with relative error guarantees, this idea removes the dependence on $m/n$ in the average complexity of BP and yields an average complexity of $\tO(1/\mu)$ or a parameterized complexity of $\tO\big(n\vpi(t)/\mu\big)$ for node $t$ (treating $\eps$ as a constant).
These complexity bounds are optimal in the sense that they match the output size in the worst case.
Additionally, in the same paper, it is shown that the complexity of BP can alternatively be bounded by $O\left(\sum_{v\in V}\vpiinv_t(v)\cdot\dout(v)/\eps\right)$ and that RBS can answer the STPPR query with absolute error bounds in $\tO\left(\sum_{v\in V}\vpiinv_t(v)\cdot\sqrt{\dout(v)}\big/\eps\right)$ time, which also outperforms BP.
These improved bounds of RBS hinge on a linear-time preprocessing phase that presorts the in-adjacency list of each node as per the out-degrees of the in-neighbors.

Liao et al.~\cite{liao2022efficient} utilize their spanning forest sampling technique to propose \textbf{BACKL}, an STPPR algorithm as a counterpart to their FORAL~\cite{liao2022efficient}.
BACKL adopts the ``BP+MC'' framework, where the MC phase is implemented by sampling random spanning forests instead of random walks.
Based on the invariant of BP, BACKL can approximate STPPR with relative error guarantees efficiently, especially for small $\alpha$.

\section{Single-Pair PPR Algorithms} \label{sec:SPPPR_algorithms}

For the SPPPR queries, Fujiwara et al.~\cite{fujiwara2012efficient} propose an exact algorithm based on \textit{node reordering}, \textit{sparse matrix representations}, and \textit{QR decomposition}; Zhao et al.~\cite{zhao2013embeddability} embed the nodes into geometric coordinate spaces to approximately answer the SPPPR query on undirected graphs.
Recently, a series of methods combine local push and MC to approximately answer the SPPPR queries.
Below, we review these methods in detail and also discuss a line of research tailored to the SNPR query.

\subsection{SPPPR Algorithms Based on Local Push and MC}

Peter Lofgren et al. propose a series of \textit{bidirectional} algorithms for the SPPPR query with relative error bounds~\cite{lofgren2014fast,lofgren2016personalized,lofgren2015bidirectional}.
First, recall that MC offers a baseline complexity of $O\left(\frac{\log(1/\delta)}{\eps^2\mu}\right)$ for this problem.
Their seminal algorithm \textbf{FAST-PPR}~\cite{lofgren2014fast} first runs BP to construct a \textit{frontier set} of the target node and then uses MC to detect the probability of reaching this frontier from the source node.
They prove that FAST-PPR achieves an average complexity (over the target node $t$) of $\tO\left(\frac{1}{\eps^2}\sqrt{\frac{m/n}{\mu}}\right)$.
In the same paper, a lower bound of $\Omega\left(\sqrt{1/\mu}\right)$ is also proved, implying that FAST-PPR's dependence on $\mu$ is optimal.

Their subsequent work \textbf{BiPPR}~\cite{lofgren2016personalized} achieves a better average complexity of $O\left(\frac{1}{\eps}\sqrt{\frac{m/n\cdot\log(1/\delta)}{\mu}}\right)$.
BiPPR also uses BP and MC but under a different analysis framework based on the invariant of BP (\Eqn~\eqref{BP_invariant}).
In a nutshell, the invariant can be rewritten as
\begin{align*}
    \vpiinv_t(s)=\vepib_t(s)+\E_{v\sim\vpi_s}\left[\vrb_t(v)\right],
\end{align*}
where the PPR vector $\vpi_s$ is regarded as a probability distribution over $V$.
Now, the expectation term can be straightforwardly estimated by running MC.

Note that FAST-PPR and BiPPR only provide average-case complexity guarantees instead of worst-case ones.
In contrast, Lofgren et al. show that on undirected graphs, their algorithm \textbf{Undirected-BiPPR}~\cite{lofgren2015bidirectional} can solve the query in $O\left(\frac{1}{\eps}\sqrt{\frac{d(t)\log(1/\delta)}{\mu}}\right)$ time.
Undirected-BiPPR combines FP and MC in a similar way as BiPPR with the aid of \Thm~\ref{thm:symmetry}.

Wang et al.~\cite{wang2016hubppr} further propose \textbf{HubPPR} as an optimized version of BiPPR.
HubPPR adopts the same bidirectional framework as BiPPR but uses two preprocessed \textit{backward oracle} and \textit{forward oracle} to speed up the BP and MC phases, resp.
Particularly, HubPPR chooses some important ``hub'' nodes and preprocesses the results of backward pushes and random walks from them so that these results can be directly used to reduce the computational cost during query time.

\subsection{Algorithms for the Single-Node PageRank Query} \label{sec:SNPR}

The SPPPR algorithms FAST-PPR~\cite{lofgren2014fast}, BiPPR~\cite{lofgren2016personalized}, and HubPPR~\cite{wang2016hubppr} can be naturally adapted for the SNPR query by modifying their MC phase from the source node.
Besides, there exists a line of work tailored to the SNPR query.
Aiming for sublinear complexities, these methods are also mainly based on local exploration and MC.

\header{Graph-Access Models.}
This line of work gives upper and lower bounds on the computational complexity or \textit{query complexity} of the SNPR query under different \textit{graph-access models}, which specify how the algorithms can access the graph through queries to a \textit{graph oracle}.
The typical setting is the \textit{arc-centric graph-access model}~\cite{bressan2018sublinear,bressan2023sublinear}, under which one can learn the in-degree/out-degree of a node or its $i$-th incoming/outgoing edge using a single query.
This model is also used by default in other works discussed in this paper.
In another \textit{node-centric graph-access model}~\cite{chen2004local,bar2008local,bressan2018sublinear,bressan2023sublinear}, one can learn the in-degree, out-degree, all in-neighbors, and all out-neighbors of a node using a single query.
Crucially, to achieve sublinear complexities, the graph oracle must additionally support a $\jump()$ query that returns a uniformly random node in $V$; otherwise, the problem cannot be answered with a sublinear query complexity in the worst case~\cite{bar2008local,bressan2013power}.

The problem of estimating single-node PageRank with sublinear complexity is first introduced in~\cite{chen2004local} by Chen et al.
Under the node-centric graph-access model, they propose several methods with a general three-phase framework: \textit{expansion}, \textit{estimation}, and \textit{iteration}.
These algorithms expand backward from the target node to obtain a subgraph, then use various heuristic approaches to estimate the PageRank values of the nodes on the boundary of the subgraph, and finally iterate over the subgraph to obtain an approximate result.

Bar-Yossef and Mashiach~\cite{bar2008local} reconsider this problem and give some lower bounds under the node-centric graph-access model.
They prove that under this model and without the $\jump()$ query, $\Omega\big(\sqrt{n}\big)$ queries are needed to estimate single-node PageRank in the worst case.
This bound holds for randomized algorithms; for deterministic algorithms, a stronger bound of $\Omega(n)$ is verified.
They show that the difficulty comes from two factors of the graph: (i) the existence of high-in-degree nodes and (ii) the slow convergence of the random walks.
These lower bounds are later further discussed in \cite{bressan2013power}.
In \cite{bar2008local}, the authors also propose an algorithm akin to those in \cite {chen2004local}, which is efficient when the graph is free from these two difficulties.
However, the existence of high-in-degree nodes in large real-world graphs makes their algorithm impractical.

\header{Sublinear-Time SNPR Algorithms.}
Bressan, Peserico, and Pretto~\cite{bressan2018sublinear,bressan2023sublinear} propose the first fully sublinear algorithms for the problem on general directed graphs.
The authors propose two novel techniques called \textit{perfect weighted estimator} and \textit{blacklisting high-degree nodes} and combine them with RPI and MC to achieve a computational complexity of $\tO\left(n^{2/3}\cdot\min\left(\Delta^{1/3},m^{1/6}\right)\right)$, where $\Delta$ is the maximum out-degree of the graph and $\text{poly}\left(\eps^{-1}\right)$ factors are omitted.
Their perfect weighted estimator has balanced coefficients and desirable concentration behavior.
It is derived as a weighted sum of several \textit{subgraph estimators}, which are similar to the estimators derived in FAST-PPR~\cite{lofgren2014fast}.
The authors also establish query complexities of $\tO\left(n^{1/2}\cdot\min\left(\Delta^{1/2},m^{1/4}\right)\right)$ under the arc-centric model and $\tO\left(\min\left(n^{1/2}\Delta^{1/2},n^{2/3}\right)\right)$ under the node-centric one.
They also give an $\Omega\left(\min\left(n^{1/2}\Delta^{1/2},n^{1/3}m^{1/3}\right)\right)$ lower bound under the arc-centric model and $\Omega\left(\min\left(n^{1/2}\Delta^{1/2},n^{2/3}\right)\right)$ under the node-centric one, implying that their algorithm is optimal under the node-centric model.

Recently, Wang and Wei~\cite{wang2023estimating} propose \textbf{SetPush} as an algorithm for the SNPR query specifically on undirected graphs.
By exploiting the symmetry property of PPR on undirected graphs (\Thm~\ref{thm:symmetry}), SetPush designs a novel push operation that combines the reverse push and the MC sampling techniques, achieving a complexity of $\tO\left(\min\big(d(t),\sqrt{m}\big)\right)$, breaking the lower bounds above for directed graphs.
Notably, SetPush does not use the $\jump()$ query.

\section{Top-$k$ PPR Algorithms} \label{sec:topk_PPR_algorithms}

There exists various algorithms devoted to the top-$k$ PPR query.
Unlike the queries above, the quality of the results for the top-$k$ PPR queries cannot be easily evaluated using previous error measures, and existing algorithms provide error guarantees from diverse aspects.
We will discuss the error guarantees explicitly for almost every algorithm.
As we did for SSPPR algorithms, we categorize top-$k$ PPR algorithms into two types based on their methods.

\subsection{Top-$k$ PPR Algorithms Based on Algebraic Techniques}

Fujiwara et al. propose a series of algorithms for exact top-$k$ PPR computation~\cite{fujiwara2012fast,fujiwara2012efficient,fujiwara2013efficient,fujiwara2013fast}.
Roughly speaking, these algorithms iteratively compute PPR estimates with upper and lower bounds and use a \textit{pruning} strategy to eliminate non-top-$k$ nodes, thereby constructing smaller subgraphs to expedite the estimation process.
Specifically, \textbf{K-dash}~\cite{fujiwara2012fast} constructs a \textit{breadth-first search tree} and exploits \textit{sparse matrix computation}, \textit{node reordering}, and \textit{LU decomposition}; \cite{fujiwara2012efficient} applies the SPPPR algorithm in \cite{fujiwara2012efficient}; \textbf{Castanet}~\cite{fujiwara2013efficient} is designed to return the exact ranking of the top-$k$ nodes among a target node set; and \textbf{F-Rank}~\cite{fujiwara2013fast} is tailored to exact top-$k$ PageRank.

Wu et al. propose another exact top-$k$ PPR algorithm called \textbf{FLoS\_RWR}~\cite{wu2014fast}.
FLoS\_RWR is derived as an extension of \textbf{FLoS\_PHP} for the computation of \textit{PHP} (\textit{penalized hitting probability}).
Using the property of PHP, FLoS\_PHP searches the graph locally and uses SPI to compute the lower and upper bounds of the PHP values for the nodes on the boundary of the explored subgraph.
By expanding the subgraph and tightening the lower and upper bounds are gradually tightened, the top-$k$ nodes can be determined.
Based on an equation between PHP and RWR, FLoS\_PHP can be modified to FLoS\_RWR without modifying the overall process.
In the paper, it is stated that the proximity values between two adjacent iterations are highly close.
Thus, the number of iterations needed to update the proximity is fairly small.

Recall that \textbf{\textsc{Chopper}}~\cite{coskun2016efficient} computes the PPR vector by running modified SPI based on Chebyshev polynomials, as discussed in \Sec~\ref{sec:adv_alg_SSPPR_PI}.
For the top-$k$ PPR query, \textsc{Chopper} takes advantage of the error bound of this Chebyshev acceleration to prune the nodes that cannot be the top-$k$ results, further reducing the number of iterations.

\subsection{Top-$k$ PPR Algorithms Based on Local Push and MC} \label{sec:adv_alg_topPPR_MC}

Sarl{\'{o}}s et al.~\cite{sarlos2006randomize} apply deterministic \textit{rounding} to the dynamic programming algorithm~\cite{jeh2003scaling} for a version of approximate top-$k$ PPR query~\cite{sarlos2006randomize} under the framework of two-phase algorithms (mentioned in Sec.~\ref{sec:adv_alg_SSPPR_MC}).
They apply rounding to the intermediate vectors of the algorithm and build a database with optimal space for this type of query.

Gupta et al.~\cite{gupta2008fast} propose the \textbf{Basic Push Algorithm} for the top-$k$ PPR query.
The number of nodes returned by this algorithm varies in a specified range.
The algorithm implements FP with a priority queue and precomputed hub vectors, obtaining lower and upper bounds for the PPR values.
Once these bounds can guarantee valid results, the algorithm terminates and returns the results.

Chakrabarti et al.~\cite{chakrabarti2011index} concentrate on building effective indices for top-$k$ PPR search so that a query can be processed in empirically constant time.
Their two solutions, \textbf{\textsc{HubRankP}} and \textbf{\textsc{HubRankD}}, are based on BCA~\cite{berkhin2006bookmark} and the decomposition theorem (\Thm~\ref{thm:decomposition}), resp., improving their early versions in \cite{gupta2008fast,pathak2008index,chakrabarti2007dynamic}.
While both algortihms achieve small index space, efficient index building and query processing, and high accuracy, \textsc{HubRankP} is considered superior to \textsc{HubRankD}.

Avrachenkov et al.~\cite{avrachenkov2011quick} utilize MC to answer the approximate top-$k$ PPR query efficiently.
They use the common MC method or ``MC complete path''~\cite{avrachenkov2007monte} to estimate the PPR vector and obtain the top-$k$ nodes.
They give approximations based on the central limit theorem to show that the algorithm achieves high efficiency if allowing a few erroneous results.

\header{Adaptations to Top-$\boldsymbol{k}$ Query.}
Several SSPPR and SPPPR algorithms mentioned before can be adapted for top-$k$ PPR queries, including BiPPR~\cite{lofgren2016personalized}, HubPPR~\cite{wang2016hubppr}, and FORA~\cite{wang2019efficient}.
BiPPR~\cite{lofgren2016personalized} considers the top-$k$ PPR query among a node set in the context of personalized search, and it applies preprocessing and \textit{hierarchical sampling} to the original BiPPR, achieving high performance along with theoretical guarantees.
HubPPR~\cite{wang2016hubppr} studies a new version of approximate top-$k$ PPR query~\cite{wang2016hubppr} among a node set.
Based on the original HubPPR, it increases the accuracy of BP and MC iteratively, and in each iteration, the lower and upper PPR bounds of all target nodes are updated, based on which some target nodes are pruned.
\textbf{FORA}~\cite{wang2019efficient} also solves the approximate top-$k$ PPR query defined in \cite{wang2016hubppr} but on the entire node set $V$.
It adopts a \textit{trial-and-error} approach to iteratively run the original FORA with different relative error thresholds, and it also computes lower and upper PPR bounds to determine when to terminate.

Wei et al.~\cite{wei2018topppr} meld FP, BP, and MC to create \textbf{TopPPR}.
Given parameters $s$, $k$, and $\rho$, TopPPR returns approximate top-$k$ nodes with the highest PPR values w.r.t. $s$ and ensures a precision of at least $\rho$.
Conceptually, TopPPR first performs FP from $s$ and simulates random walks from the nodes with nonzero residues, as FORA does; it then runs BP \textit{adaptively} from some candidate nodes to determine the top-$k$ nodes.
TopPPR estimates the PPR values through a combination of the invariants of both FP and BP, and it maintains and expands a set containing the nodes that are certain to be among the top-$k$.
TopPPR runs in $O\left((m+n\log n)\big/\sqrt{\mathrm{gap}_{\rho}}\right)$ time and specifically in $O\left(k^{1/4}n^{3/4}\log n\big/\sqrt{\mathrm{gap}_{\rho}}\right)$ time on power-law graphs, where $\mathrm{gap}_{\rho}$ is the difference between the $\lceil\rho k\rceil$-th and the $(k+1)$-th largest PPR value w.r.t. $s$.
Notably, the complexity is sublinear in $n$ on power-law graphs if $k=o(n)$.

\section{Algorithms for Other PPR Queries} \label{sec:other_queries}

This section discusses PPR algorithms tailored to some specific queries.
These queries are relatively less studied but useful for some particular applications.

\header{Subgraph PageRank Query.}
Davis and Dhillon~\cite{davis2006estimating} study the problem of locally estimating the PageRank values in a given subgraph.
They propose to expand the given subgraph by greedily selecting a series of nodes on the frontier and to utilize \textit{stochastic complementation} to compute the PageRank values on the expanded subgraph.
Their approach runs in linear time w.r.t. the number of nodes of the given subgraph, assuming that the number of performed iterations is constant.

\header{Subgraph PageRank Ranking.}
Bressan and Pretto~\cite{bressan2011local} theoretically study the problem of determining the PageRank ranking of $k$ given nodes.
They show that, in the worst case, a correct algorithm must explore $\Omega(n)$ nodes if it is deterministic and $\Omega\left(\sqrt{kn}\right)$ nodes if randomized.
The later work~\cite{bressan2013power} by Bressan et al. further shows that Las Vegas algorithms must explore $n-o(n)$ nodes and that MC algorithms also need to explore $\Omega(n)$ nodes if the $\jump()$ query (see \Sec~\ref{sec:SNPR}) is not allowed.
However, if the $\jump()$ query is allowed, the lower bound of MC algorithms degrades to $\Omega\left(n^{2/3}\right)$.
For this case, they propose \textbf{SampleRank} that achieves sublinear complexity when the PageRank values in question are not too small.

\header{Significant PageRank Query.}
Borgs, Brautbar, Chayes, and Teng~\cite{borgs2012sublinear,borgs2014multiscale} focus on the \textit{Significant PageRank} problem: given $\Delta$ and $c>1$, with high probability compute a set that contains all nodes with PageRank value at least $\Delta$ and no node smaller than $\Delta/c$.
Based on an abstraction of the problem to a basic matrix problem, their approach utilizes two novel components, \textit{multiscale matrix sampling} and \textit{robust approximation of PPR}, to solve this problem in $\tO(1/\Delta)$ time.
Their robust approximation of PPR aims to estimate PPR values within a relative error plus an absolute error, and their process resembles the standard MC method.

\header{Reverse Top-$\boldsymbol{k}$ PPR Query.}
Yu et al.~\cite{yu2014reverse} study the \textit{reverse top-$k$ PPR query}, which asks for all nodes that have $t$ in their top-$k$ PPR node sets, given the target node $t$ and an integer $k$.
Their proposed algorithm comprises an indexing technique and an online search process.
In the preprocessing phase, BCA~\cite{berkhin2006bookmark} is invoked to produce lower bounds of the top PPR values, which are stored compactly as the index.
In the online querying phase, the index is used in cooperation with RPI to prune or confirm the candidate nodes, where the computational overhead can be reduced, and the index can be refined for better future performance.

\header{Heavy Hitters Query.}
Wang and Tao~\cite{wang2018efficient} delve into identifying approximate PPR \textit{heavy hitters} and propose \textbf{BLOG} for this purpose.
Concretely, given two node sets $S$ and $T$, BLOG approximately finds all pairs $(s,t)$ such that $s\in S$, $t\in T$, and $\vpi_s(t)$ is above a specified fraction of $\vpi(t)$.
BLOG primarily combines MC and BP to determine whether each pair of $(s,t)$ satisfies the property, and it incorporates a novel technique to circumvent high-in-degree nodes when performing BP.
Moreover, BLOG shares some computational results across different node pairs to reduce redundancy.
Also, BLOG leverages a new \textit{logarithmic bucketing} technique that partitions the target nodes into groups and applies different parameter settings for each group.
Overall, BLOG achieves superior theoretical complexity and empirical performance.

\header{Batch One-Hop PPR Query.}
Luo et al.~\cite{luo2019efficient,luo2019baton} propose \textbf{Baton} for a new type of PPR query known as the \textit{batch one-hop PPR query}: given a node set $S\subseteq V$, Baton provides an estimate of $\vpi_s(v)$ for each pair of $s\in S$ and $v\in\Nout(s)$ with relative error guarantees.
Baton follows the ``FP+MC'' framework of FORA~\cite{wang2019efficient} but adopts novel techniques and analysis tailored to this query.
In particular, Baton adopts distinct stopping criteria for the FP process and reuses the generated random walks.
Its analysis exploits a lower bound for one-hop PPR and the concept of ``generic Baton.''
It is shown that Baton runs in $O\big(\sum_{s\in S}\dout(s)\log(1/\delta)/\eps^2\big)$ expected time, which is nearly optimal.
The authors also consider applying a \textit{vertex-cover-based optimization} on undirected graphs and multicore parallelization to Baton.

\section{PPR Computation in Special Settings} \label{sec:special}

This section reviews several algorithms for computing PPR in some special settings.
Sarkar and Moore~\cite{sarkar2010fast} consider finding top nodes with the largest degree-normalized PPR values on large disk-resident undirected graphs, and their methods are based on a graph clustering algorithm and the dynamic programming algorithm in~\cite{sarlos2006randomize}.
Yang~\cite{yang2022efficient} computes PPR and its extensions on bipartite graphs using modifications of PI and FP.
Below, we particularly introduce the challenges and some algorithms for the settings of dynamic graphs, parallel/distributed settings, weighted graphs, and graph streams.

\subsection{For Dynamic Graphs}

\header{Settings and Challenges.}
Although most relevant works focus on PPR computation on \textit{static graphs}, many real-world graphs for practical applications, such as social networks, are constantly evolving.
This necessitates the development of PPR algorithms tailored to \textit{dynamic graphs}, where the graph undergoes continuous updates, including edge insertions and deletions.
Efficiently maintaining PPR on dynamic graphs poses some unique challenges.
Firstly, running linear-time PPR algorithms upon each update is prohibitive due to the frequent updates.
Secondly, several efficient algorithms introduced before rely on precomputed indices to speed up the query process, but they cannot be directly applied to dynamic graphs.
In fact, updating these indices over time can be rather costly.
Finally, ensuring accuracy guarantees for evolving graphs introduces technical difficulties and probably requires more sophisticated approaches.
To address these challenges, several PPR algorithms for dynamic graphs have been proposed.

Bahmani et al.~\cite{bahmani2010fast} apply MC to maintain PageRank values or top-$k$ PPR nodes for a dynamic graph, where $m$ edges arrive in an equally random order to a graph with $n$ nodes, and the edges can also be deleted randomly over time.
This setting is called the \textit{random arrival model} and is adopted in some later works~\cite{zhang2016approximate,mo2021agenda,hou2023personalized}.
For PageRank values, their method is based on the ``MC complete path'' method~\cite{avrachenkov2007monte}.
Roughly speaking, they perform several random walks from each node, store the entire paths, and use the number of occurrences of the nodes to estimate their PageRank values.
Then, for an edge insertion, only a small portion of the stored paths need to be updated in expectation.
They demonstrate that the expected complexity to maintain all PageRank values at all times with constant error is $O(n\ln m)$.
This method also supports a random edge removal with $O(n/m)$ expected work.
Also, a similar approach works for top-$k$ PPR maintenance under more complicated models and assumptions.

Unlike other studies for dynamic graphs, Bahmani et al.~\cite{bahmani2012pagerank} consider the case when the algorithm is oblivious to the changes in the graph.
They propose effective crawling strategies for periodically probing the nodes in the graph to detect the changes and update the PageRank vector.

Yu and Lin~\cite{yu2013irwr} propose \textbf{IRWR} for dynamically computing exact SSPPR.
They first derive a formula for the new PPR vector $\vpi_s'$ after inserting or deleting an edge $(u,v)$ to/from the graph.
This formula expresses $\vpi_s'$ as a linear combination of $\vpi_u$ and $\vpi_v$ before the operation.
Based on this, given the original graph, its PPR matrix (computed using other algorithms), the source node $s$, and a batch of edge insertion/deletion operations, IRWR can incrementally update the PPR vectors of $s$ and other relevant nodes, and thus obtain the resultant PPR vector of $s$.

A later work of Yu and McCann~\cite{yu2016random} considers maintaining the PPR matrix over dynamic graphs exactly, and their solution is called \textbf{Inc-R}.
By characterizing the changes to $\bPi$ as the outer product of two vectors and skipping unnecessary computation, Inc-R can handle a single edge insertion/deletion in time proportional to the number of affected entries in $\bPi$.

Ohsaka et al.~\cite{ohsaka2015efficient} propose to adopt FP for maintaining dynamic SSPPR with absolute error guarantees on an evolving graph.
Their algorithm \textbf{\textsc{TrackingPPR}} maintains the reserve and residue vectors dynamically by possibly performing pushes when the graph is updated.
For a single edge insertion or deletion, their algorithm only needs to perform amortized $O(1/\eps)$ pushes, and for $m$ randomly inserted edges, the total number of pushes is $O(\log m/\eps)$ in expectation.

Zhang et al.~\cite{zhang2016approximate} adopt FP and BP to dynamic graphs, yielding efficient algorithms for dynamically maintaining approximate PPR and inverse PPR vectors, resp.
The algorithms are dubbed \textbf{\textsc{LazyForward}} and \textbf{\textsc{LazyReverse}}.
Conceptually, for every edge insertion/deletion, their algorithms first maintain the invariant between the reserve and residue vectors and then perform push operations to restore accuracy if needed.
For PPR, \textsc{LazyForward} maintains degree-normalized error on an undirected graph, with the expected complexity of $O(1/\eps)$ plus $O(1)$ per update, for a uniformly random source node $s$.
For inverse PPR, \textsc{LazyReverse} maintains absolute error on a directed graph in the random arrival model~\cite{bahmani2010fast} or on an arbitrary undirected graph, with an expected complexity of $O(m/n/\eps)$ plus $O(1)$ per update, for a uniformly random target node $t$.

Yoon et al.~\cite{yoon2018fast} devise accurate dynamic SSPPR algorithms \textbf{OSP} and \textbf{OSP-T} based on CPI.
For an edge insertion/deletion, OSP uses CPI to compute the difference between the old PPR vector and the new one by expressing it as a power series of the difference of the transition matrix.
OSP-T further introduces a trade-off between accuracy and efficiency by limiting the number of iterations in CPI.

Mo and Luo~\cite{mo2021agenda,mo2023single} present \textbf{Agenda}, an efficient algorithm for maintaining SSPPR with relative error guarantees on dynamic graphs.
Agenda is designed to be ``robust,'' in the sense that it sustains efficiency regardless of whether the ratio of updates to queries is high or low.
Following the framework of FORA+~\cite{wang2019efficient}, Agenda stores precomputed random walks as indices, which are then combined with FP's results at query time.
Crucially, Agenda runs BP to compute a quantitative measure of the index inaccuracy and adopts a \textit{lazy-update} mechanism to reduce the overhead of updating the indices over time.
Theoretically, under the random arrival model~\cite{bahmani2010fast}, the query complexity of Agenda matches that of FORA+, and the expected fraction of updated indices is also bounded.

A subsequent work, \textbf{FIRM} by Hou et al.~\cite{hou2023personalized}, also follows the framework of FORA+~\cite{wang2019efficient} but imposes an \textit{eager-update} strategy to update the indices of random walks.
Notably, FIRM maintains auxiliary data structures to support updating the indices in expected $O(1)$ time under the random arrival model~\cite{bahmani2010fast}.
Furthermore, FIRM can use a novel \textit{sampling scheme} to reduce the space consumption of these data structures while preserving the expected $O(1)$ update complexity.

\subsection{In Parallel/Distributed Settings} \label{sec:distributed}

\header{Settings and Challenges.}
The enormous sizes of current graphs have made it imperative to develop efficient algorithms for computing PPR in \textit{parallel} or \textit{distributed} settings, where the algorithm can allocate the computation workload to multiple processors or machines for high performance and scalability.
The detailed settings vary across different works, and we will introduce them briefly below.
The core challenge under these settings lies in effectively managing the computational tasks across multiple processors or machines, each handling a different part of the graph.
Meanwhile, the algorithms need to minimize the number of computational rounds and the communication cost while ensuring desirable convergence rates or error guarantees.
Particularly, as it is nontrivial to parallelize the basic techniques of PI and FP, more sophisticated techniques are required to adapt existing methods for PPR computation to parallel or distributed settings.

Some early works~\cite{sankaralingam2003distributed,shi2003distributed,wang2004computing,parreira2006efficient} aim at computing PageRank on a \textit{decentralized peer-to-peer network}, where each peer computes PageRank values individually using iterative methods and communicates with others to refine and combine the results.
Another early work~\cite{zhu2005distributed} uses \textit{divide-and-conquer} to compute local PageRank vectors and then update them through communication with a coordinator.

Bahmani et al.~\cite{bahmani2011fast} propose \textbf{DoublingPPR} for approximately computing the PPR matrix under the \textit{MapReduce} model.
Their method is based on MC, and its basic idea is to generate random walks of a given length for all nodes efficiently using MapReduce.
To achieve this, DoublingPPR merges short segments of random walks to obtain longer random walks in a \textit{doubling} manner.

Mitliagkas et al. propose \textbf{\textsc{FrogWild}}~\cite{mitliagkas2015frogwild} for approximating top-$k$ PageRank using the \textit{GraphLab} framework.
\textsc{FrogWild} generates parallel random walks to estimate high PageRank values, and it adopts \textit{partial synchronization} to reduce the network traffic.
Although the generated random walks are no longer independent, the authors give analysis for their correlations and bound the approximation errors.

Liu et al.~\cite{liu2016powerwalk} propose \textbf{PowerWalk} to approximate PPR vectors under a distributed framework.
It simulates full-path random walks as offline indexing, and for online query, it computes linear combinations of these indices in a \textit{vertex-centric decomposition} manner to obtain approximate PPR vectors.
PowerWalk strikes an adaptive trade-off between offline preprocessing and online query based on the memory budget.

Guo et al.~\cite{guo2017distributed} introduce \textbf{GPA} and \textbf{HGPA} for computing exact SSPPR on a \textit{coordinator-based share-nothing distributed computing platform}.
The salient idea of GPA is to \textit{partition} the graph into disjoint subgraphs and distribute the computation accordingly, and HGPA further splits the subgraphs recursively to form a \textit{tree-structured hierarchy}.
Based on precomputed data, GPA and HGPA can allocate the computation evenly to separate machines, guaranteeing that only one time of communication between each machine and the coordinator is needed to obtain the aggregated PPR vector at query time.

Guo et al.~\cite{guo2017parallel} migrate the algorithm for dynamic SSPPR in~\cite{zhang2016approximate} to parallel settings of GPUs or multicore CPUs.
Their algorithm is a \textit{batch-processing} method with novel techniques for updating the reserve and residue vectors in parallel, leading to reduced synchronization cost and a higher level of parallelism.
While the algorithm's complexity matches the original sequential one~\cite{zhang2016approximate}, its specially designed optimizations yield extensive practical speedups.

Shi et al.~\cite{shi2019realtime} propose \textbf{kPAR} for answering the approximate top-$k$ PPR query defined in \cite{wang2016hubppr} using GPUs on a single commodity machine.
To exploit the power of GPU, kPAR utilizes novel techniques called \textit{adaptive forward push} and \textit{inverted random walks} and performs careful system engineering specific to GPUs.
kPAR runs in $O\left(n+\frac{1}{\eps}\sqrt{\frac{m\log(1/\delta)}{\mu}}\right)$ time and can answer the query in real-time on billion-scale graphs.

Lin~\cite{lin2019distributed} exploits the \textit{parallel pipeline framework} to compute MC approximation of the PPR matrix with relative error bounds on large edge-weighted graphs.
Their \textbf{DistPPR} utilizes three carefully designed techniques to efficiently generate many random walks from each node.
In particular, these techniques help to handle nodes with large degrees, precompute walk segments for nodes with small degrees, and optimize the workload for each pipeline to reduce the overhead.

Wang et al.~\cite{wang2019parallelizing} endeavor to parallelize FORA~\cite{wang2019efficient} on shared-memory multicore systems under the \textit{work-depth model}.
Their work-efficient approach \textbf{PAFO} achieves linear speedup in the number of processors.
PAFO exploits various novel techniques, including efficient maintenance of the active nodes in FP, \textit{cache-aware scheduling}, and an \textit{integer-counting-based method} for generating random walks.

Hou et al.~\cite{hou2021massively} propose \textbf{Delta-Push} for the SSPPR and top-$k$ PPR queries under the \textit{Massively Parallel Computing (MPC) model}.
Delta-Push follows the framework of FORA+~\cite{wang2019efficient} and combines a redesigned \textit{parallel push} algorithm and MC in a nontrivial way.
They show that Delta-Push reduces the number of rounds to $O\left( \log\frac{n^2\log n}{\eps^{2}m} \right)$ with a load of $O(m/p)$, where $\eps$ is the relative error bound and $p$ is the number of executors.

\header{Other Theoretical Results.}
Some additional theoretical works focus on the round complexity and the communication bandwidth of computing PageRank under distributed settings.
Most of these works use a \textit{stitching} technique to generate long random walks by combining short random walks in a doubling manner.
In \cite{sarma2015fast}, the authors consider the \textit{round complexity} of computing PageRank under a kind of \textit{congested clique model}, where only limited messages can be sent over each edge in each round.
They devise a special method to simulate random walks in parallel, which avoids communication congestion.
Their algorithms take $O(\log n)$ rounds on directed graphs and $O\left( \sqrt{\log n}\right)$ rounds on undirected graphs, with edge bandwidth bounded by $O(\log n)$ and $O\left(\log^{3}n\right)$ bits, resp.
On undirected graphs, the round complexity of $O\left( \sqrt{\log n}\right)$ is later improved to $O(\log\log n)$ by Luo~\cite{luo2019distributed}, and the bandwidth is also improved in a subsequent work by Luo~\cite{luo2020improved}.
The \textbf{Multiple-Round Radar-Push} algorithm proposed in \cite{luo2019distributed} uses a novel approach to generate a number of low-variance random walks from each node, and \cite{luo2020improved} further combines this algorithm with FP to reduce the communication cost.

Some other works investigate the MPC model instead of the congested clique model.
\cite{lacki2020walking} proves the following bounds on the round complexity for computing PageRank under the MPC model: $O(\log\log n)$ for undirected graphs (with $\tO(m)$ total space) and $\tO\left(\log^2\log n\right)$ for directed graphs (with $\tO\left(m+n^{1+o(1)}\right)$ total space).
Kapralov et al.~\cite{kapralov2021efficient} further study the complexity of computing PPR vectors under the MPC model and give some more involved bounds.

\subsection{For Weighted Graphs}

\header{Settings and Challenges.}
This subsection considers computing PPR on graphs with weighted edges.
On such weighted graphs, when an $\alpha$-discounted random walk proceeds, each out-neighbor is selected with a probability proportional to the weight of the connecting edge instead of with an equal probability.
Some applications of PPR on weighted graphs are introduced in \cite{xie2015edge,wang2022edge}.
While some previously introduced PPR algorithms are adaptable for weighted graphs, specially designed algorithms can outperform these adaptations by a large margin.
However, there is a limited body of research for this specific purpose.
We introduce two works for computing SSPPR on weighted graphs below: the first employs algebraic techniques, and the second utilizes the local-push method.

The first empirically efficient algorithm tailored to weighted graphs is proposed by Xie et al.~\cite{xie2015edge}.
Their method relies on \textit{model reduction}, a general approach for approximating solutions to large linear systems.
Note that on weighted graphs, the PPR vector still satisfies a linear equation similar to \Eqn~\eqref{eqn:PPR_def}, with $\P$ replaced by the corresponding weighted transition matrix.
Based on model reduction, their algorithm can reduce this linear system to a low dimension, solve the reduced system, and reconstruct the approximate PPR vector.
They also apply optimizations for parameterizations and cost-accuracy trade-offs to further improve their algorithm's empirical performance.
However, their algorithm is still costly as it needs to process the large matrices of the graph.

Wang et al. propose \textbf{EdgePush}~\cite{wang2022edge} to optimize FP on weighted graphs.
EdgePush decomposes the atomic node-level push operation in FP into separate edge-level pushes, assigning different pushing thresholds for each edge.
With a delicately designed data structure, each push operation can be done in amortized $O(1)$ time.
This \textit{edge-based push} strategy endows EdgePush with superior complexities over FP, especially on graphs where the edge weight distribution is imbalanced.
In such cases, the complexity bound is improved by a factor up to $\Theta(n)$.
The authors give detailed complexity bounds for both $\ell_1$-error and degree-normalized absolute error guarantees.

\subsection{For Graph Streams}

\header{Settings and Challenges.}
This subsection reviews algorithms for computing PageRank when the edges of the underlying graph are presented as a \textit{stream}.
The objective under this streaming model is to compute PageRank using minimal space and few passes over the stream.
Under these constraints, even sampling a random walk becomes a complex task, and there are only a few relevant theoretical studies.

Sarma, Gollapudi, and Panigrahy~\cite{sarma2008estimating,sarma2011estimating} consider estimating the PageRank vector on graph streams.
Based on a new streaming algorithm for estimating the \textit{mixing time} of random walks, their algorithm only requires $o(n)$ space and $\tO\big(\sqrt{M}\big)$ passes to compute PageRank values above $M/n$, where $M$ denotes the mixing time.
Kallaugher, Kapralov, and Price~\cite{kallaugher2022simulating} investigate the setting where the edges are given in a uniformly random order.
They design a \textit{single-pass} algorithm for generating nearly independent random walks and apply it to estimate the sum of PageRank values for a node set with absolute error guarantees using sublinear space.

\section{Research Challenges and Directions} \label{sec:directions}

\Tbls~\ref{tab:summary} and \ref{tab:summary_special} summarize the representative PPR algorithms under the typical setting and special settings, resp.
Despite the extensive body of research, numerous unresolved issues persist in the realm of efficient PPR computation from both theoretical and practical perspectives.

\begin{itemize}[leftmargin=*]
    \item From the \textit{theoretical} point of view, the major concern is the sublinear complexities of various PPR approximation problems.
    For example, for the basic techniques, can one optimize FP or BP by modifying their push strategies or termination conditions?
    As the error bound of FP and the worst-case complexity of BP on general directed graphs are still unclear, FP and BP's properties may not have been fully understood.
    Also, the gaps between the upper and lower bounds for the SPPPR query~\cite{lofgren2014fast,lofgren2015bidirectional} and the SNPR query~\cite{bressan2018sublinear,bressan2023sublinear}, along with the problem of solving $\ell_1$-regularized PageRank~\cite{fountoulakis2022open}, remain as open problems.
    It is also desirable to establish lower bounds for other PPR approximation problems and narrow the gaps between the upper and lower bounds.
    Finally, it is interesting to further investigate the complexities for undirected graphs, dynamic graphs, weighted graphs, etc.
    \item From the \textit{practical} point of view, enhancing the efficiency and scalability of PPR computation is still challenging.
    The vast scale of real-world graphs greatly limits the applicability of classic iterative PPR algorithms, necessitating approximate algorithms that harness the power of parallel or distributed computing.
    An important but unsolved problem is to approximately compute or factorize the whole PPR matrix of large graphs efficiently~\cite{yin2019scalable,bojchevski2020scaling,tsitsulin2021frede}.
    Moreover, there is a notable absence of a standardized benchmark or an extensive experimental survey for evaluating the practical performance of existing algorithms.
    As current algorithms cover various settings and traditional worst-case complexity analysis cannot adequately reflect their empirical performance, such a benchmark or experimental survey is meaningful and essential.
    Finally, some interesting future directions include using machine learning to accelerate PPR computation and using PPR to develop differentially private algorithms for downstream graph learning tasks~\cite{epasto2022differentially}.
\end{itemize}

\begin{table*}[!t]
    \caption{Summary of Representative PPR Algorithms under the Typical Setting}
    \label{tab:summary}
    \centering
    \begin{tabular}{lllll}
        \toprule
        \textbf{Query} & \textbf{Algorithm} & \textbf{Main Techniques/Tools} & \textbf{Error Bound} & \textbf{Notes} \\
        \midrule
        \multirow{18}{*}{SSPPR}& \cite{sarlos2006randomize} & dynamic programming, Count-Min Sketch & absolute & \tabincell{l}{two-phase algorithm with \\ optimal database space} \\
        & FastPPV~\cite{zhu2013incremental} & prioritizing paths, scheduled estimation & $\ell_1$ \\
        & \cite{maehara2014computing} & \tabincell{l}{core-tree decomposition, preconditioning} & $\ell_1$ \\
        & BEAR~\cite{jung2016random} & block elimination, reordering, LU decomposition & exact & has approximate version \\
        & \textsc{BePI}~\cite{jung2017bepi} & block elimination, sparsifying, preconditioning & $\ell_2$-error & improves over BEAR \\
        & \textsc{Chopper}~\cite{coskun2016efficient} & SPI, Chebyshev polynomials & not given \\
        & TPA~\cite{yoon2018tpa} & dividing the summation of PPR into three parts & $\ell_1$ \\
        & \cite{chen2023accelerating} & momentum-based acceleration methods & $\ell_1$ & \tabincell{l}{on undirected graphs, \\ $1/\sqrt{\alpha}$ dependence on $\alpha$} \\
        \cline{2-5}
        \addlinespace[1pt]
        & \cite{chung2010sharp} & FP with dynamic threshold & same as FP & exponential convergence \\
        & PowerPush~\cite{wu2021unifying} & unifying FIFO-FP and PI & $\ell_1$ & exponential convergence \\
        & FORA, FORA+~\cite{wang2019efficient} & FP+MC, indexing & relative \\
        & ResAcc~\cite{lin2020index} & FP+MC, residue accumulation & relative & index-free \\
        & SpeedPPR~\cite{wu2021unifying} & PowerPush+MC, indexing & relative \\
        & FORAL~\cite{liao2022efficient} & FP+spanning forest sampling & see \cite{liao2022efficient} & robust for small $\alpha$ \\
        & \cite{liao2023efficient} & PI, progressive and spanning forest sampling & relative \\
        & \cite{epasto2022differentially} & FP with residue restriction & absolute & bounded sensitivity \\
        \midrule
        \multirow{3}{*}{STPPR} & \cite{lofgren2013personalized} & BP with priority queue & absolute & exponential convergence \\
        & RBS~\cite{wang2020personalized} & randomized BP, presorting & absolute/relative & optimal for relative error \\
        & BACKL~\cite{liao2022efficient} & BP+spanning forest sampling & relative & robust for small $\alpha$ \\
        \midrule
        \multirow{6}{*}{SPPPR}& FAST-PPR~\cite{lofgren2014fast} & BP+MC, frontier set & relative & optimal dependence on $\mu$ \\
        & BiPPR~\cite{lofgren2016personalized} & BP+MC, bidirectional estimator & relative & improves over FAST-PPR \\
        & Undirected-BiPPR~\cite{lofgren2015bidirectional} & FP+MC, symmetry of PPR & relative & on undirected graphs \\
        & HubPPR~\cite{wang2016hubppr} & BP+MC, indexing & relative & improves over BiPPR \\
        & \cite{bressan2018sublinear,bressan2023sublinear} & local exploration, MC, perfect estimator, blacklist & relative & SNPR, sublinear complexity \\
        & SetPush~\cite{wang2023estimating} & BP, MC, symmetry of PPR & relative & SNPR on undirected graphs \\
        \midrule
        \multirow{9}{*}{top-$k$ PPR} & \cite{fujiwara2012fast,fujiwara2012efficient,fujiwara2013efficient,fujiwara2013fast} & computing lower/upper bounds, pruning & exact & \tabincell{l}{\cite{fujiwara2013efficient} returns exact ranking, \\ \cite{fujiwara2013fast} is for top-$k$ PageRank} \\
        & FLoS\_RWR~\cite{wu2014fast} & local search, SPI on subgraphs & exact \\
        & \textsc{Chopper}~\cite{coskun2016efficient} & Chebyshev polynomials, SPI, pruning & exact \\
        \cline{2-5}
        \addlinespace[1pt]
        & \cite{sarlos2006randomize} & dynamic programming, rounding & see \cite{sarlos2006randomize} & \tabincell{l}{two-phase algorithm with \\ optimal database space} \\
        & \textsc{HubRank}~\cite{chakrabarti2011index} & BCA, indexing & not given \\
        & \cite{lofgren2016personalized,wang2016hubppr,wang2019efficient} & FP, BP, MC, hierarchical sampling & see \cite{lofgren2016personalized,wang2016hubppr} & adaptations for top-$k$ query \\
    	& TopPPR~\cite{wei2018topppr} & FP+BP+MC, adaptive BP & $\rho$-precision & \tabincell{l}{sublinear complexity in $n$ \\ on power-law graphs} \\
        \bottomrule
    \end{tabular}
\end{table*}

\begin{table*}[!t]
    \caption{Summary of Representative PPR Algorithms under Special Settings}
    \label{tab:summary_special}
    \centering
    \begin{tabular}{lllll}
        \toprule
        \textbf{Setting} & \textbf{Algorithm} & \textbf{Main Techniques/Tools} & \textbf{Notes} \\
        \midrule
        \multirow{10}{*}{Dynamic Graphs} & \cite{bahmani2010fast} & MC, update random walks & for PageRank or top-$k$ PPR, random arrival model \\
        & \cite{bahmani2012pagerank} & effective crawling strategies & for PageRank, oblivious to the changes in the graph \\
        & IRWR~\cite{yu2013irwr} & algebraically update PPR vector & for exact SSPPR \\
        & Inc-R~\cite{yu2016random} & algebraically update PPR matrix & for exact PPR matrix \\
        & \textsc{TrackingPPR}~\cite{ohsaka2015efficient} & FP, pushes as per updates & for SSPPR with absolute error bounds \\
        & \cite{zhang2016approximate} & FP, BP, maintaining invariants & for SSPPR or STPPR, similar error bounds to FP/BP \\
        & OSP, OSP-T~\cite{yoon2018fast} & CPI, computing the difference & for SSPPR, OSP-T balances accuracy and efficiency \\
        & Agenda~\cite{mo2021agenda,mo2023single} & FP+MC, indexing, lazy-update & for SSPPR with relative error bounds, workload robustness \\
        & FIRM~\cite{hou2023personalized} & FP+MC, indexing, eager-update & for SSPPR with relative error bounds \\
        \midrule
        \multirow{9}{*}{Parallel/Distributed} & DoublingPPR~\cite{bahmani2011fast} & MC, doubling & for PPR matrix, MapReduce model \\
        & \textsc{FrogWild}~\cite{mitliagkas2015frogwild} & MC, partial synchronization & for approximate top-$k$ PageRank, GraphLab framework \\
        & PowerWalk~\cite{liu2016powerwalk} & MC, vertex-centric decomposition & for approximate SSPPR, balances indexing and query \\
        & GPA, HGPA~\cite{guo2017distributed} & partitioning, tree structure & for exact SSPPR, coordinator-based platform \\
        & \cite{guo2017parallel} & \cite{zhang2016approximate}, batch-processing & for SSPPR, on GPUs or multicore CPUs \\
        & kPAR~\cite{shi2019realtime} & adaptive FP, inverted random walks & for approximate top-$k$ PPR, on GPUs \\
        & DistPPR~\cite{lin2019distributed} & MC, workload optimization & for PPR matrix on weighted graphs, pipeline framework \\
        & PAFO~\cite{wang2019parallelizing} & FP+MC, cache-aware scheduling & for SSPPR, parallelizes FORA, work-depth model \\
        & Delta-Push~\cite{hou2023personalized} & FP+MC, parallel push & for SSPPR or top-$k$ PPR, MPC model \\
        \midrule
        \multirow{2}{*}{Weighted Graphs} & \cite{xie2015edge} & model reduction & for SSPPR, based on algebraic techniques \\
        & EdgePush~\cite{wang2022edge} & edge-based FP & for SSPPR, superior complexity for imbalanced weights \\
        \midrule
        \multirow{2}{*}{Graph Streams} & \cite{sarma2008estimating,sarma2011estimating} & estimating mixing time & estimates PageRank values above a threshold \\
        & \cite{kallaugher2022simulating} & generating random walks & estimates PageRank values, single-pass, on random stream \\
        \bottomrule
    \end{tabular}
\end{table*}

\section{Conclusions} \label{sec:conclusion}

This survey studies efficient PPR computation and elaborates on five basic techniques: the Monte Carlo method, Power Iteration, Forward Push, Reverse Power Iteration, and Backward Push.
We then conduct an extensive survey of recent PPR algorithms from an algorithmic perspective, summarizing their rationales and contributions.
Our taxonomy is based on the query types, including Single-Source PPR, Single-Target PPR, Single-Pair PPR, and Top-$k$ PPR queries.
In addition, we review several algorithms tailored to special settings, such as on dynamic graphs and in parallel/distributed environments.
We also give high-level comparisons of PPR algorithms based on their techniques and point out future research directions, aiming to inspire and guide subsequent advancements in the field of PPR computation.

\section*{Acknowledgments}

This research was supported by NSFC of China (No. U2241212, No. 61932001, No. U2001212, No. U1936205), Beijing Natural Science Foundation (No. 4222028), Beijing Outstanding Young Scientist Program No.BJJWZYJH012019100020098, Hong Kong RGC GRF (No. 14217322), Hong Kong RGC CRF (No. C4158-20G), Hong Kong ITC ITF (No. MRP/071/20X), CUHK Direct Grant (No. 4055181), and Engineering Research Center of Next-Generation Intelligent Search and Recommendation, Ministry of Education. We thank the anonymous reviewers for their valuable comments.

\bibliographystyle{IEEEtran}
\bibliography{paper}

\begin{thebibliography}{100}
\providecommand{\url}[1]{#1}
\csname url@samestyle\endcsname
\providecommand{\newblock}{\relax}
\providecommand{\bibinfo}[2]{#2}
\providecommand{\BIBentrySTDinterwordspacing}{\spaceskip=0pt\relax}
\providecommand{\BIBentryALTinterwordstretchfactor}{4}
\providecommand{\BIBentryALTinterwordspacing}{\spaceskip=\fontdimen2\font plus
\BIBentryALTinterwordstretchfactor\fontdimen3\font minus
  \fontdimen4\font\relax}
\providecommand{\BIBforeignlanguage}[2]{{%
\expandafter\ifx\csname l@#1\endcsname\relax
\typeout{** WARNING: IEEEtran.bst: No hyphenation pattern has been}%
\typeout{** loaded for the language `#1'. Using the pattern for}%
\typeout{** the default language instead.}%
\else
\language=\csname l@#1\endcsname
\fi
#2}}
\providecommand{\BIBdecl}{\relax}
\BIBdecl

\bibitem{brin1998anatomy}
S.~Brin and L.~Page, ``The anatomy of a large-scale hypertextual web search
  engine,'' \emph{Comput. Netw.}, vol.~30, no. 1-7, pp. 107--117, 1998.

\bibitem{wang2018efficient}
S.~Wang and Y.~Tao, ``Efficient algorithms for finding approximate heavy
  hitters in personalized pageranks,'' in \emph{Proc. ACM SIGMOD Int. Conf.
  Manage. Data}, 2018, pp. 1113--1127.

\bibitem{wu2008top}
X.~Wu, V.~Kumar, J.~R. Quinlan, J.~Ghosh, Q.~Yang, H.~Motoda, G.~J. McLachlan,
  A.~F.~M. Ng, B.~Liu, P.~S. Yu, Z.~Zhou, M.~S. Steinbach, D.~J. Hand, and
  D.~Steinberg, ``Top 10 algorithms in data mining,'' \emph{Knowl. Inf. Syst.},
  vol.~14, no.~1, pp. 1--37, 2008.

\bibitem{gleich2015pagerank}
D.~F. Gleich, ``Pagerank beyond the web,'' \emph{SIAM Rev.}, vol.~57, no.~3,
  pp. 321--363, 2015.

\bibitem{andersen2006local}
R.~Andersen, F.~R.~K. Chung, and K.~J. Lang, ``Local graph partitioning using
  pagerank vectors,'' in \emph{Proc. 47th Annu. IEEE Symp. Found. Comput.
  Sci.}, 2006, pp. 475--486.

\bibitem{andersen2007pagerank}
------, ``Using pagerank to locally partition a graph,'' \emph{Internet Math.},
  vol.~4, no.~1, pp. 35--64, 2007.

\bibitem{yin2017local}
H.~Yin, A.~R. Benson, J.~Leskovec, and D.~F. Gleich, ``Local higher-order graph
  clustering,'' in \emph{Proc. 23rd ACM SIGKDD Int. Conf. Knowl. Discovery Data
  Mining}, 2017, pp. 555--564.

\bibitem{yang2019efficient}
R.~Yang, X.~Xiao, Z.~Wei, S.~S. Bhowmick, J.~Zhao, and R.~Li, ``Efficient
  estimation of heat kernel pagerank for local clustering,'' in \emph{Proc. ACM
  SIGMOD Int. Conf. Manage. Data}, 2019, pp. 1339--1356.

\bibitem{fountoulakis2019variational}
K.~Fountoulakis, F.~Roosta{-}Khorasani, J.~Shun, X.~Cheng, and M.~W. Mahoney,
  ``Variational perspective on local graph clustering,'' \emph{Math. Program.},
  vol. 174, no. 1-2, pp. 553--573, 2019.

\bibitem{chang2022community}
Y.~Chang, H.~Ma, L.~Chang, and Z.~Li, ``Community detection with attributed
  random walk via seed replacement,'' \emph{Frontiers Comput. Sci.}, vol.~16,
  no.~5, p. 165324, 2022.

\bibitem{yuan2024index}
Z.~Yuan, Z.~Wei, F.~Lv, and J.~Wen, ``Index-free triangle-based graph local
  clustering,'' \emph{Frontiers Comput. Sci.}, vol.~18, no.~3, p. 183404, 2024.

\bibitem{ou2016asymmetric}
M.~Ou, P.~Cui, J.~Pei, Z.~Zhang, and W.~Zhu, ``Asymmetric transitivity
  preserving graph embedding,'' in \emph{Proc. 22nd ACM SIGKDD Int. Conf.
  Knowl. Discovery Data Mining}, 2016, pp. 1105--1114.

\bibitem{zhou2017scalable}
C.~Zhou, Y.~Liu, X.~Liu, Z.~Liu, and J.~Gao, ``Scalable graph embedding for
  asymmetric proximity,'' in \emph{Proc. 31st AAAI Conf. Artif. Intell.}, 2017,
  pp. 2942--2948.

\bibitem{tsitsulin2018verse}
A.~Tsitsulin, D.~Mottin, P.~Karras, and E.~M{\"{u}}ller, ``Verse: Versatile
  graph embeddings from similarity measures,'' in \emph{Proc. Int. Conf. World
  Wide Web}, 2018, pp. 539--548.

\bibitem{yin2019scalable}
Y.~Yin and Z.~Wei, ``Scalable graph embeddings via sparse transpose
  proximities,'' in \emph{Proc. 25th ACM SIGKDD Int. Conf. Knowl. Discovery
  Data Mining}, 2019, pp. 1429--1437.

\bibitem{yang2020homegeneous}
R.~Yang, J.~Shi, X.~Xiao, Y.~Yang, and S.~S. Bhowmick, ``Homogeneous network
  embedding for massive graphs via reweighted personalized pagerank,''
  \emph{Proc. VLDB Endowment}, vol.~13, no.~5, pp. 670--683, 2020.

\bibitem{tsitsulin2021frede}
A.~Tsitsulin, M.~Munkhoeva, D.~Mottin, P.~Karras, I.~V. Oseledets, and
  E.~M{\"{u}}ller, ``Frede: Anytime graph embeddings,'' \emph{Proc. VLDB
  Endowment}, vol.~14, no.~6, pp. 1102--1110, 2021.

\bibitem{klicpera2018predict}
J.~Klicpera, A.~Bojchevski, and S.~G{\"{u}}nnemann, ``Predict then propagate:
  Graph neural networks meet personalized pagerank,'' in \emph{Proc. 7th Int.
  Conf. Learn. Representations}, 2019.

\bibitem{bojchevski2020scaling}
A.~Bojchevski, J.~Klicpera, B.~Perozzi, A.~Kapoor, M.~Blais,
  B.~R{\'{o}}zemberczki, M.~Lukasik, and S.~G{\"{u}}nnemann, ``Scaling graph
  neural networks with approximate pagerank,'' in \emph{Proc. 26th ACM SIGKDD
  Int. Conf. Knowl. Discovery Data Mining}, 2020, pp. 2464--2473.

\bibitem{chen2020scalable}
M.~Chen, Z.~Wei, B.~Ding, Y.~Li, Y.~Yuan, X.~Du, and J.~Wen, ``Scalable graph
  neural networks via bidirectional propagation,'' in \emph{Proc. Annu. Conf.
  Neural Inf. Process. Syst.}, 2020.

\bibitem{wang2021approximate}
H.~Wang, M.~He, Z.~Wei, S.~Wang, Y.~Yuan, X.~Du, and J.~Wen, ``Approximate
  graph propagation,'' in \emph{Proc. 27th ACM SIGKDD Int. Conf. Knowl.
  Discovery Data Mining}, 2021, pp. 1686--1696.

\bibitem{zhang2016accuracy}
Y.~Zhang, C.~Li, C.~Xie, and H.~Chen, ``Accuracy estimation of link-based
  similarity measures and its application,'' \emph{Frontiers Comput. Sci.},
  vol.~10, no.~1, pp. 113--123, 2016.

\bibitem{epasto2022differentially}
A.~Epasto, V.~Mirrokni, B.~Perozzi, A.~Tsitsulin, and P.~Zhong,
  ``Differentially private graph learning via sensitivity-bounded personalized
  pagerank,'' in \emph{Proc. Annu. Conf. Neural Inf. Process. Syst.}, 2022.

\bibitem{yang2022efficient}
R.~Yang, ``Efficient and effective similarity search over bipartite graphs,''
  in \emph{ACM Web Conf.}, 2022, pp. 308--318.

\bibitem{zhang2023effective}
S.~Zhang, R.~Yang, X.~Xiao, X.~Yan, and B.~Tang, ``Effective and efficient
  pagerank-based positioning for graph visualization,'' \emph{Proc. ACM Manage.
  Data}, vol.~1, no.~1, pp. 76:1--76:27, 2023.

\bibitem{moler2002world}
C.~Moler, ``The world’s largest matrix computation,'' \emph{Matlab News
  Notes}, pp. 12--13, 2002.

\bibitem{berkhin2005survey}
P.~Berkhin, ``Survey: A survey on pagerank computing,'' \emph{Internet Math.},
  vol.~2, no.~1, pp. 73--120, 2005.

\bibitem{chung2014brief}
F.~Chung, ``A brief survey of pagerank algorithms,'' \emph{IEEE Trans. Netw.
  Sci. Eng.}, vol.~1, no.~1, pp. 38--42, 2014.

\bibitem{park2019survey}
S.~Park, W.~Lee, B.~Choe, and S.~Lee, ``A survey on personalized pagerank
  computation algorithms,'' \emph{IEEE Access}, vol.~7, pp. 163\,049--163\,062,
  2019.

\bibitem{avrachenkov2007monte}
K.~Avrachenkov, N.~Litvak, D.~Nemirovsky, and N.~Osipova, ``Monte carlo methods
  in pagerank computation: When one iteration is sufficient,'' \emph{SIAM J.
  Numer. Anal.}, vol.~45, no.~2, pp. 890--904, 2007.

\bibitem{bianchini2005inside}
M.~Bianchini, M.~Gori, and F.~Scarselli, ``Inside pagerank,'' \emph{ACM Trans.
  Internet Technol.}, vol.~5, no.~1, pp. 92--128, 2005.

\bibitem{jeh2003scaling}
G.~Jeh and J.~Widom, ``Scaling personalized web search,'' in \emph{Proc. Int.
  Conf. World Wide Web}, 2003, pp. 271--279.

\bibitem{avrachenkov2013choice}
K.~Avrachenkov, P.~Gon{\c{c}}alves, and M.~Sokol, ``On the choice of kernel and
  labelled data in semi-supervised learning methods,'' in \emph{Proc. 10th Int.
  Workshop Algorithms Models Web Graph}, vol. 8305, 2013, pp. 56--67.

\bibitem{langville2004deeper}
A.~N. Langville and C.~D. Meyer, ``Survey: Deeper inside pagerank,''
  \emph{Internet Math.}, vol.~1, no.~3, pp. 335--380, 2003.

\bibitem{fogaras2005towards}
D.~Fogaras, B.~R{\'{a}}cz, K.~Csalog{\'{a}}ny, and T.~Sarl{\'{o}}s, ``Towards
  scaling fully personalized pagerank: Algorithms, lower bounds, and
  experiments,'' \emph{Internet Math.}, vol.~2, no.~3, pp. 333--358, 2005.

\bibitem{andersen2007local}
R.~Andersen, C.~Borgs, J.~T. Chayes, J.~E. Hopcroft, V.~S. Mirrokni, and
  S.~Teng, ``Local computation of pagerank contributions,'' in \emph{Proc. 5th
  Int. Workshop Algorithms Models Web Graph}, vol. 4863, 2007, pp. 150--165.

\bibitem{andersen2008local}
------, ``Local computation of pagerank contributions,'' \emph{Internet Math.},
  vol.~5, no.~1, pp. 23--45, 2008.

\bibitem{wei2018topppr}
Z.~Wei, X.~He, X.~Xiao, S.~Wang, S.~Shang, and J.~Wen, ``Topppr: Top-k
  personalized pagerank queries with precision guarantees on large graphs,'' in
  \emph{Proc. ACM SIGMOD Int. Conf. Manage. Data}, 2018, pp. 441--456.

\bibitem{liao2022efficient}
M.~Liao, R.~Li, Q.~Dai, and G.~Wang, ``Efficient personalized pagerank
  computation: A spanning forests sampling based approach,'' in \emph{Proc. ACM
  SIGMOD Int. Conf. Manage. Data}, 2022, pp. 2048--2061.

\bibitem{saad2003iterative}
Y.~Saad, \emph{Iterative methods for sparse linear systems}.\hskip 1em plus
  0.5em minus 0.4em\relax SIAM, 2003.

\bibitem{yoon2018tpa}
M.~Yoon, J.~Jung, and U.~Kang, ``Tpa: Fast, scalable, and accurate method for
  approximate random walk with restart on billion scale graphs,'' in
  \emph{Proc. 34th Int. Conf. Data Eng.}, 2018, pp. 1132--1143.

\bibitem{chen2023accelerating}
Z.~Chen, X.~Guo, B.~Zhou, D.~Yang, and S.~Skiena, ``Accelerating personalized
  pagerank vector computation,'' in \emph{Proc. 29th ACM SIGKDD Int. Conf.
  Knowl. Discovery Data Mining}, 2023, pp. 262--273.

\bibitem{wu2021unifying}
H.~Wu, J.~Gan, Z.~Wei, and R.~Zhang, ``Unifying the global and local
  approaches: An efficient power iteration with forward push,'' in \emph{Proc.
  ACM SIGMOD Int. Conf. Manage. Data}, 2021, pp. 1996--2008.

\bibitem{kamvar2003extrapolation}
S.~D. Kamvar, T.~H. Haveliwala, C.~D. Manning, and G.~H. Golub, ``Extrapolation
  methods for accelerating pagerank computations,'' in \emph{Proc. Int. Conf.
  World Wide Web}, 2003, pp. 261--270.

\bibitem{broder2004efficient}
A.~Z. Broder, R.~Lempel, F.~Maghoul, and J.~O. Pedersen, ``Efficient pagerank
  approximation via graph aggregation,'' in \emph{Proc. Int. Conf. World Wide
  Web}, 2004, pp. 484--485.

\bibitem{kamvar2004adaptive}
S.~Kamvar, T.~Haveliwala, and G.~Golub, ``Adaptive methods for the computation
  of pagerank,'' \emph{Linear Algebra Appl.}, vol. 386, pp. 51--65, 2004.

\bibitem{wu2010arnoldi}
G.~Wu and Y.~Wei, ``Arnoldi versus gmres for computing pagerank: A theoretical
  contribution to google's pagerank problem,'' \emph{ACM Trans. Inf. Syst.},
  vol.~28, no.~3, pp. 11:1--11:28, 2010.

\bibitem{tong2006fast}
H.~Tong, C.~Faloutsos, and J.~Pan, ``Fast random walk with restart and its
  applications,'' in \emph{Proc. 6th Int. Conf. Data Mining}, 2006, pp.
  613--622.

\bibitem{gleich2007approximating}
D.~F. Gleich and M.~Polito, ``Approximating personalized pagerank with minimal
  use of web graph data,'' \emph{Internet Math.}, vol.~3, no.~3, pp. 257--294,
  2007.

\bibitem{sarlos2006randomize}
T.~Sarl{\'{o}}s, A.~A. Bencz{\'{u}}r, K.~Csalog{\'{a}}ny, D.~Fogaras, and
  B.~R{\'{a}}cz, ``To randomize or not to randomize: space optimal summaries
  for hyperlink analysis,'' in \emph{Proc. Int. Conf. World Wide Web}, 2006,
  pp. 297--306.

\bibitem{zhu2013incremental}
F.~Zhu, Y.~Fang, K.~C. Chang, and J.~Ying, ``Incremental and accuracy-aware
  personalized pagerank through scheduled approximation,'' \emph{Proc. VLDB
  Endowment}, vol.~6, no.~6, pp. 481--492, 2013.

\bibitem{maehara2014computing}
T.~Maehara, T.~Akiba, Y.~Iwata, and K.~Kawarabayashi, ``Computing personalized
  pagerank quickly by exploiting graph structures,'' \emph{Proc. VLDB
  Endowment}, vol.~7, no.~12, pp. 1023--1034, 2014.

\bibitem{shin2015bear}
K.~Shin, J.~Jung, L.~Sael, and U.~Kang, ``Bear: Block elimination approach for
  random walk with restart on large graphs,'' in \emph{Proc. ACM SIGMOD Int.
  Conf. Manage. Data}, 2015, pp. 1571--1585.

\bibitem{jung2016random}
J.~Jung, K.~Shin, L.~Sael, and U.~Kang, ``Random walk with restart on large
  graphs using block elimination,'' \emph{ACM Trans. Database Syst.}, vol.~41,
  no.~2, pp. 12:1--12:43, 2016.

\bibitem{jung2017bepi}
J.~Jung, N.~Park, L.~Sael, and U.~Kang, ``Bepi: Fast and memory-efficient
  method for billion-scale random walk with restart,'' in \emph{Proc. ACM
  SIGMOD Int. Conf. Manage. Data}, 2017, pp. 789--804.

\bibitem{coskun2016efficient}
M.~Co{\c{s}}kun, A.~Grama, and M.~Koyut{\"{u}}rk, ``Efficient processing of
  network proximity queries via chebyshev acceleration,'' in \emph{Proc. 22nd
  ACM SIGKDD Int. Conf. Knowl. Discovery Data Mining}, 2016, pp. 1515--1524.

\bibitem{berkhin2006bookmark}
P.~Berkhin, ``Bookmark-coloring algorithm for personalized pagerank
  computing,'' \emph{Internet Math.}, vol.~3, no.~1, pp. 41--62, 2006.

\bibitem{chakrabarti2007dynamic}
S.~Chakrabarti, ``Dynamic personalized pagerank in entity-relation graphs,'' in
  \emph{Proc. Int. Conf. World Wide Web}, 2007, pp. 571--580.

\bibitem{chung2010sharp}
F.~C. Graham and W.~Zhao, ``A sharp pagerank algorithm with applications to
  edge ranking and graph sparsification,'' in \emph{Proc. 7th Int. Workshop
  Algorithms Models Web Graph}, vol. 6516, 2010, pp. 2--14.

\bibitem{wang2017fora}
S.~Wang, R.~Yang, X.~Xiao, Z.~Wei, and Y.~Yang, ``Fora: Simple and effective
  approximate single-source personalized pagerank,'' in \emph{Proc. 23rd ACM
  SIGKDD Int. Conf. Knowl. Discovery Data Mining}, 2017, pp. 505--514.

\bibitem{wang2019efficient}
S.~Wang, R.~Yang, R.~Wang, X.~Xiao, Z.~Wei, W.~Lin, Y.~Yang, and N.~Tang,
  ``Efficient algorithms for approximate single-source personalized pagerank
  queries,'' \emph{ACM Trans. Database Syst.}, vol.~44, no.~4, pp. 18:1--18:37,
  2019.

\bibitem{luo2019efficient}
S.~Luo, X.~Xiao, W.~Lin, and B.~Kao, ``Efficient batch one-hop personalized
  pageranks,'' in \emph{Proc. 35th Int. Conf. Data Eng.}, 2019, pp. 1562--1565.

\bibitem{luo2019baton}
------, ``Baton: Batch one-hop personalized pageranks with efficiency and
  accuracy,'' \emph{IEEE Trans. Knowl. Data Eng.}, vol.~32, no.~10, pp.
  1897--1908, 2020.

\bibitem{wang2019parallelizing}
R.~Wang, S.~Wang, and X.~Zhou, ``Parallelizing approximate single-source
  personalized pagerank queries on shared memory,'' \emph{VLDB J.}, vol.~28,
  no.~6, pp. 923--940, 2019.

\bibitem{lin2020index}
D.~Lin, R.~C. Wong, M.~Xie, and V.~J. Wei, ``Index-free approach with
  theoretical guarantee for efficient random walk with restart query,'' in
  \emph{Proc. 36th Int. Conf. Data Eng.}, 2020, pp. 913--924.

\bibitem{luo2020improved}
S.~Luo, ``Improved communication cost in distributed pagerank computation - a
  theoretical study,'' in \emph{Proc. 37th Int. Conf. Mach. Learn.}, vol. 119,
  2020, pp. 6459--6467.

\bibitem{hou2021massively}
G.~Hou, X.~Chen, S.~Wang, and Z.~Wei, ``Massively parallel algorithms for
  personalized pagerank,'' \emph{Proc. VLDB Endowment}, vol.~14, no.~9, pp.
  1668--1680, 2021.

\bibitem{mo2021agenda}
D.~Mo and S.~Luo, ``Agenda: Robust personalized pageranks in evolving graphs,''
  in \emph{Proc. 30th ACM Int. Conf. Inf. Knowl. Manage.}, 2021, pp.
  1315--1324.

\bibitem{liao2023efficient}
M.~Liao, R.~Li, Q.~Dai, H.~Chen, H.~Qin, and G.~Wang, ``Efficient personalized
  pagerank computation: The power of variance-reduced monte carlo approaches,''
  \emph{Proc. ACM Manage. Data}, vol.~1, no.~2, pp. 160:1--160:26, 2023.

\bibitem{hou2023personalized}
G.~Hou, Q.~Guo, F.~Zhang, S.~Wang, and Z.~Wei, ``Personalized pagerank on
  evolving graphs with an incremental index-update scheme,'' \emph{Proc. ACM
  Manage. Data}, vol.~1, no.~1, pp. 25:1--25:26, 2023.

\bibitem{fountoulakis2022open}
K.~Fountoulakis and S.~Yang, ``Open problem: Running time complexity of
  accelerated $\ell_1$-regularized pagerank,'' in \emph{Annu. Conf. Learn.
  Theory}, 2022, pp. 5630--5632.

\bibitem{martinez2023accelerated}
D.~Mart{\'{\i}}nez{-}Rubio, E.~S. Wirth, and S.~Pokutta, ``Accelerated and
  sparse algorithms for approximate personalized pagerank and beyond,'' in
  \emph{Annu. Conf. Learn. Theory}, vol. 195, 2023, pp. 2852--2876.

\bibitem{lofgren2013personalized}
P.~Lofgren and A.~Goel, ``Personalized pagerank to a target node,''
  \emph{CoRR}, vol. abs/1304.4658, 2013.

\bibitem{wang2020personalized}
H.~Wang, Z.~Wei, J.~Gan, S.~Wang, and Z.~Huang, ``Personalized pagerank to a
  target node, revisited,'' in \emph{Proc. 26th ACM SIGKDD Int. Conf. Knowl.
  Discovery Data Mining}, 2020, pp. 657--667.

\bibitem{fujiwara2012efficient}
Y.~Fujiwara, M.~Nakatsuji, T.~Yamamuro, H.~Shiokawa, and M.~Onizuka,
  ``Efficient personalized pagerank with accuracy assurance,'' in \emph{Proc.
  18th ACM SIGKDD Int. Conf. Knowl. Discovery Data Mining}, 2012, pp. 15--23.

\bibitem{zhao2013embeddability}
X.~Zhao, A.~Chang, A.~D. Sarma, H.~Zheng, and B.~Y. Zhao, ``On the
  embeddability of random walk distances,'' \emph{Proc. VLDB Endowment},
  vol.~6, no.~14, pp. 1690--1701, 2013.

\bibitem{lofgren2014fast}
P.~Lofgren, S.~Banerjee, A.~Goel, and S.~Comandur, ``Fast-ppr: scaling
  personalized pagerank estimation for large graphs,'' in \emph{Proc. 20th ACM
  SIGKDD Int. Conf. Knowl. Discovery Data Mining}, 2014, pp. 1436--1445.

\bibitem{lofgren2016personalized}
P.~Lofgren, S.~Banerjee, and A.~Goel, ``Personalized pagerank estimation and
  search: A bidirectional approach,'' in \emph{Proc. 9th ACM Int. Conf. Web
  Search Data Mining}, 2016, pp. 163--172.

\bibitem{lofgren2015bidirectional}
------, ``Bidirectional pagerank estimation: From average-case to worst-case,''
  in \emph{Proc. 12th Int. Workshop Algorithms Models Web Graph}, vol. 9479,
  2015, pp. 164--176.

\bibitem{wang2016hubppr}
S.~Wang, Y.~Tang, X.~Xiao, Y.~Yang, and Z.~Li, ``Hubppr: Effective indexing for
  approximate personalized pagerank,'' \emph{Proc. VLDB Endowment}, vol.~10,
  no.~3, pp. 205--216, 2016.

\bibitem{bressan2018sublinear}
M.~Bressan, E.~Peserico, and L.~Pretto, ``Sublinear algorithms for local graph
  centrality estimation,'' in \emph{Proc. 59th Annu. IEEE Symp. Found. Comput.
  Sci.}, 2018, pp. 709--718.

\bibitem{bressan2023sublinear}
------, ``Sublinear algorithms for local graph-centrality estimation,''
  \emph{SIAM J. Comput.}, vol.~52, no.~4, pp. 968--1008, 2023.

\bibitem{chen2004local}
Y.~Chen, Q.~Gan, and T.~Suel, ``Local methods for estimating pagerank values,''
  in \emph{Proc. 13th ACM Int. Conf. Inf. Knowl. Manage.}, 2004, pp. 381--389.

\bibitem{bar2008local}
Z.~Bar{-}Yossef and L.~Mashiach, ``Local approximation of pagerank and reverse
  pagerank,'' in \emph{Proc. 17th ACM Int. Conf. Inf. Knowl. Manage.}, 2008,
  pp. 279--288.

\bibitem{bressan2013power}
M.~Bressan, E.~Peserico, and L.~Pretto, ``The power of local information in
  pagerank,'' in \emph{Proc. Int. Conf. World Wide Web}, 2013, pp. 179--180.

\bibitem{wang2023estimating}
H.~Wang and Z.~Wei, ``Estimating single-node pagerank in
  $\widetilde{O}\left(\min\big\{d_t,\sqrt{m}\big\}\right)$ time,'' \emph{Proc.
  VLDB Endowment}, vol.~16, no.~11, pp. 2949--2961, 2023.

\bibitem{fujiwara2012fast}
Y.~Fujiwara, M.~Nakatsuji, M.~Onizuka, and M.~Kitsuregawa, ``Fast and exact
  top-k search for random walk with restart,'' \emph{Proc. VLDB Endowment},
  vol.~5, no.~5, pp. 442--453, 2012.

\bibitem{fujiwara2013efficient}
Y.~Fujiwara, M.~Nakatsuji, H.~Shiokawa, T.~Mishima, and M.~Onizuka, ``Efficient
  ad-hoc search for personalized pagerank,'' in \emph{Proc. ACM SIGMOD Int.
  Conf. Manage. Data}, 2013, pp. 445--456.

\bibitem{fujiwara2013fast}
------, ``Fast and exact top-k algorithm for pagerank,'' in \emph{Proc. 27th
  AAAI Conf. Artif. Intell.}, 2013, pp. 1106--1112.

\bibitem{wu2014fast}
Y.~Wu, R.~Jin, and X.~Zhang, ``Fast and unified local search for random walk
  based k-nearest-neighbor query in large graphs,'' in \emph{Proc. ACM SIGMOD
  Int. Conf. Manage. Data}, 2014, pp. 1139--1150.

\bibitem{gupta2008fast}
M.~S. Gupta, A.~Pathak, and S.~Chakrabarti, ``Fast algorithms for topk
  personalized pagerank queries,'' in \emph{Proc. Int. Conf. World Wide Web},
  2008, pp. 1225--1226.

\bibitem{chakrabarti2011index}
S.~Chakrabarti, A.~Pathak, and M.~Gupta, ``Index design and query processing
  for graph conductance search,'' \emph{VLDB J.}, vol.~20, no.~3, pp. 445--470,
  2011.

\bibitem{pathak2008index}
A.~Pathak, S.~Chakrabarti, and M.~S. Gupta, ``Index design for dynamic
  personalized pagerank,'' in \emph{Proc. 24th Int. Conf. Data Eng.}, 2008, pp.
  1489--1491.

\bibitem{avrachenkov2011quick}
K.~Avrachenkov, N.~Litvak, D.~Nemirovsky, E.~Smirnova, and M.~Sokol, ``Quick
  detection of top-k personalized pagerank lists,'' in \emph{Proc. 8th Int.
  Workshop Algorithms Models Web Graph}, vol. 6732, 2011, pp. 50--61.

\bibitem{davis2006estimating}
J.~V. Davis and I.~S. Dhillon, ``Estimating the global pagerank of web
  communities,'' in \emph{Proc. 12th ACM SIGKDD Int. Conf. Knowl. Discovery
  Data Mining}, 2006, pp. 116--125.

\bibitem{bressan2011local}
M.~Bressan and L.~Pretto, ``Local computation of pagerank: the ranking side,''
  in \emph{Proc. 20th ACM Int. Conf. Inf. Knowl. Manage.}, 2011, pp. 631--640.

\bibitem{borgs2012sublinear}
C.~Borgs, M.~Brautbar, J.~T. Chayes, and S.~Teng, ``A sublinear time algorithm
  for pagerank computations,'' in \emph{Proc. 9th Int. Workshop Algorithms
  Models Web Graph}, vol. 7323, 2012, pp. 41--53.

\bibitem{borgs2014multiscale}
------, ``Multiscale matrix sampling and sublinear-time pagerank computation,''
  \emph{Internet Math.}, vol.~10, no. 1-2, pp. 20--48, 2014.

\bibitem{yu2014reverse}
A.~W. Yu, N.~Mamoulis, and H.~Su, ``Reverse top-k search using random walk with
  restart,'' \emph{Proc. VLDB Endowment}, vol.~7, no.~5, pp. 401--412, 2014.

\bibitem{sarkar2010fast}
P.~Sarkar and A.~W. Moore, ``Fast nearest-neighbor search in disk-resident
  graphs,'' in \emph{Proc. 16th ACM SIGKDD Int. Conf. Knowl. Discovery Data
  Mining}, 2010, pp. 513--522.

\bibitem{bahmani2010fast}
B.~Bahmani, A.~Chowdhury, and A.~Goel, ``Fast incremental and personalized
  pagerank,'' \emph{Proc. VLDB Endowment}, vol.~4, no.~3, pp. 173--184, 2010.

\bibitem{zhang2016approximate}
H.~Zhang, P.~Lofgren, and A.~Goel, ``Approximate personalized pagerank on
  dynamic graphs,'' in \emph{Proc. 22nd ACM SIGKDD Int. Conf. Knowl. Discovery
  Data Mining}, 2016, pp. 1315--1324.

\bibitem{bahmani2012pagerank}
B.~Bahmani, R.~Kumar, M.~Mahdian, and E.~Upfal, ``Pagerank on an evolving
  graph,'' in \emph{Proc. 18th ACM SIGKDD Int. Conf. Knowl. Discovery Data
  Mining}, 2012, pp. 24--32.

\bibitem{yu2013irwr}
W.~Yu and X.~Lin, ``Irwr: incremental random walk with restart,'' in
  \emph{Proc. 36th ACM SIGIR Int. Conf. Res. Develop. Inf. Retrieval}, 2013,
  pp. 1017--1020.

\bibitem{yu2016random}
W.~Yu and J.~A. McCann, ``Random walk with restart over dynamic graphs,'' in
  \emph{Proc. 16th Int. Conf. Data Mining}, 2016, pp. 589--598.

\bibitem{ohsaka2015efficient}
N.~Ohsaka, T.~Maehara, and K.~Kawarabayashi, ``Efficient pagerank tracking in
  evolving networks,'' in \emph{Proc. 21st ACM SIGKDD Int. Conf. Knowl.
  Discovery Data Mining}, 2015, pp. 875--884.

\bibitem{yoon2018fast}
M.~Yoon, W.~Jin, and U.~Kang, ``Fast and accurate random walk with restart on
  dynamic graphs with guarantees,'' in \emph{Proc. Int. Conf. World Wide Web},
  2018, pp. 409--418.

\bibitem{mo2023single}
D.~Mo and S.~Luo, ``Single-source personalized pageranks with workload
  robustness,'' \emph{IEEE Trans. Knowl. Data Eng.}, vol.~35, no.~6, pp.
  6320--6334, 2023.

\bibitem{sankaralingam2003distributed}
K.~Sankaralingam, S.~Sethumadhavan, and J.~C. Browne, ``Distributed pagerank
  for p2p systems,'' in \emph{12th Int. Symp. High-Perform. Distrib. Comput.},
  2003, pp. 58--69.

\bibitem{shi2003distributed}
S.~Shi, J.~Yu, G.~Yang, and D.~Wang, ``Distributed page ranking in structured
  p2p networks,'' in \emph{32nd Int. Conf. Parallel Process.}, 2003, pp.
  179--186.

\bibitem{wang2004computing}
Y.~Wang and D.~J. DeWitt, ``Computing pagerank in a distributed internet search
  engine system,'' in \emph{Proc. 30th Int. Conf. Very Large Data Bases}, 2004,
  pp. 420--431.

\bibitem{parreira2006efficient}
J.~X. Parreira, D.~Donato, S.~Michel, and G.~Weikum, ``Efficient and
  decentralized pagerank approximation in a peer-to-peer web search network,''
  in \emph{Proc. 32nd Int. Conf. Very Large Data Bases}, 2006, pp. 415--426.

\bibitem{zhu2005distributed}
Y.~Zhu, S.~Ye, and X.~Li, ``Distributed pagerank computation based on iterative
  aggregation-disaggregation methods,'' in \emph{Proc. 14th ACM Int. Conf. Inf.
  Knowl. Manage.}, 2005, pp. 578--585.

\bibitem{bahmani2011fast}
B.~Bahmani, K.~Chakrabarti, and D.~Xin, ``Fast personalized pagerank on
  mapreduce,'' in \emph{Proc. ACM SIGMOD Int. Conf. Manage. Data}, 2011, pp.
  973--984.

\bibitem{mitliagkas2015frogwild}
I.~Mitliagkas, M.~Borokhovich, A.~G. Dimakis, and C.~Caramanis, ``Frogwild! -
  fast pagerank approximations on graph engines,'' \emph{Proc. VLDB Endowment},
  vol.~8, no.~8, pp. 874--885, 2015.

\bibitem{liu2016powerwalk}
Q.~Liu, Z.~Li, J.~C.~S. Lui, and J.~Cheng, ``Powerwalk: Scalable personalized
  pagerank via random walks with vertex-centric decomposition,'' in \emph{Proc.
  25th ACM Int. Conf. Inf. Knowl. Manage.}, 2016, pp. 195--204.

\bibitem{guo2017distributed}
T.~Guo, X.~Cao, G.~Cong, J.~Lu, and X.~Lin, ``Distributed algorithms on exact
  personalized pagerank,'' in \emph{Proc. ACM SIGMOD Int. Conf. Manage. Data},
  2017, pp. 479--494.

\bibitem{guo2017parallel}
W.~Guo, Y.~Li, M.~Sha, and K.~Tan, ``Parallel personalized pagerank on dynamic
  graphs,'' \emph{Proc. VLDB Endowment}, vol.~11, no.~1, pp. 93--106, 2017.

\bibitem{shi2019realtime}
J.~Shi, R.~Yang, T.~Jin, X.~Xiao, and Y.~Yang, ``Realtime top-k personalized
  pagerank over large graphs on gpus,'' \emph{Proc. VLDB Endowment}, vol.~13,
  no.~1, pp. 15--28, 2019.

\bibitem{lin2019distributed}
W.~Lin, ``Distributed algorithms for fully personalized pagerank on large
  graphs,'' in \emph{Proc. Int. Conf. World Wide Web}, 2019, pp. 1084--1094.

\bibitem{sarma2015fast}
A.~D. Sarma, A.~R. Molla, G.~Pandurangan, and E.~Upfal, ``Fast distributed
  pagerank computation,'' \emph{Theor. Comput. Sci.}, vol. 561, pp. 113--121,
  2015.

\bibitem{luo2019distributed}
S.~Luo, ``Distributed pagerank computation: An improved theoretical study,'' in
  \emph{Proc. 33rd AAAI Conf. Artif. Intell.}, 2019, pp. 4496--4503.

\bibitem{lacki2020walking}
J.~{\L}{\k{a}}cki, S.~Mitrovi{\'c}, K.~Onak, and P.~Sankowski, ``Walking
  randomly, massively, and efficiently,'' in \emph{Proc. 52nd Annu. ACM SIGACT
  Symp. Theory Comput.}, 2020, pp. 364--377.

\bibitem{kapralov2021efficient}
M.~Kapralov, S.~Lattanzi, N.~Nouri, and J.~Tardos, ``Efficient and local
  parallel random walks,'' in \emph{Proc. Annu. Conf. Neural Inf. Process.
  Syst.}, 2021, pp. 21\,375--21\,387.

\bibitem{xie2015edge}
W.~Xie, D.~Bindel, A.~J. Demers, and J.~Gehrke, ``Edge-weighted personalized
  pagerank: Breaking {A} decade-old performance barrier,'' in \emph{Proc. 21st
  ACM SIGKDD Int. Conf. Knowl. Discovery Data Mining}, 2015, pp. 1325--1334.

\bibitem{wang2022edge}
H.~Wang, Z.~Wei, J.~Gan, Y.~Yuan, X.~Du, and J.~Wen, ``Edge-based local push
  for personalized pagerank,'' \emph{Proc. VLDB Endowment}, vol.~15, no.~7, pp.
  1376--1389, 2022.

\bibitem{sarma2008estimating}
A.~D. Sarma, S.~Gollapudi, and R.~Panigrahy, ``Estimating pagerank on graph
  streams,'' in \emph{Proc. 27th ACM SIGMOD-SIGACT-SIGART Symp. Princ. Database
  Syst.}, 2008, pp. 69--78.

\bibitem{sarma2011estimating}
------, ``Estimating pagerank on graph streams,'' \emph{J. ACM}, vol.~58,
  no.~3, pp. 13:1--13:19, 2011.

\bibitem{kallaugher2022simulating}
J.~Kallaugher, M.~Kapralov, and E.~Price, ``Simulating random walks in random
  streams,'' in \emph{Proc. ACM-SIAM Symp. Discrete Algorithms}, 2022, pp.
  3091--3126.

\end{thebibliography}

\newpage
 
\begin{IEEEbiography}[{\includegraphics[width=1in,height=1.25in,clip,keepaspectratio]{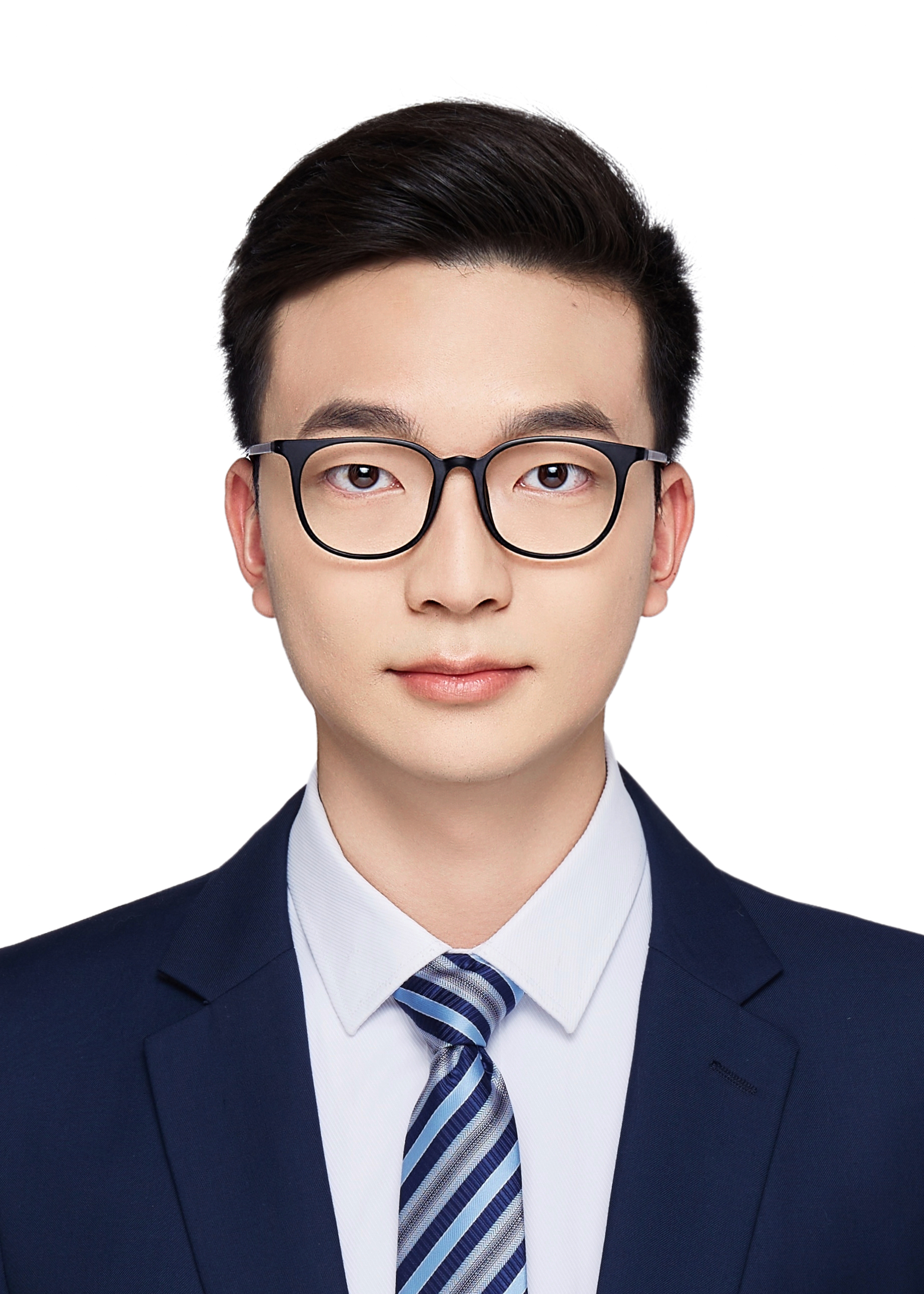}}]{Mingji Yang}
is currently pursuing his Ph.D. degree at Gaoling School of Artificial Intelligence, Renmin University of China.
He received his B.E. degree in Computer Science and Technology in 2022 from School of Information, Renmin University of China.
His research interests include graph algorithms and graph mining.
\end{IEEEbiography}

\begin{IEEEbiography}[{\includegraphics[width=1in,height=1.25in,trim=400 150 400 250,clip,keepaspectratio]{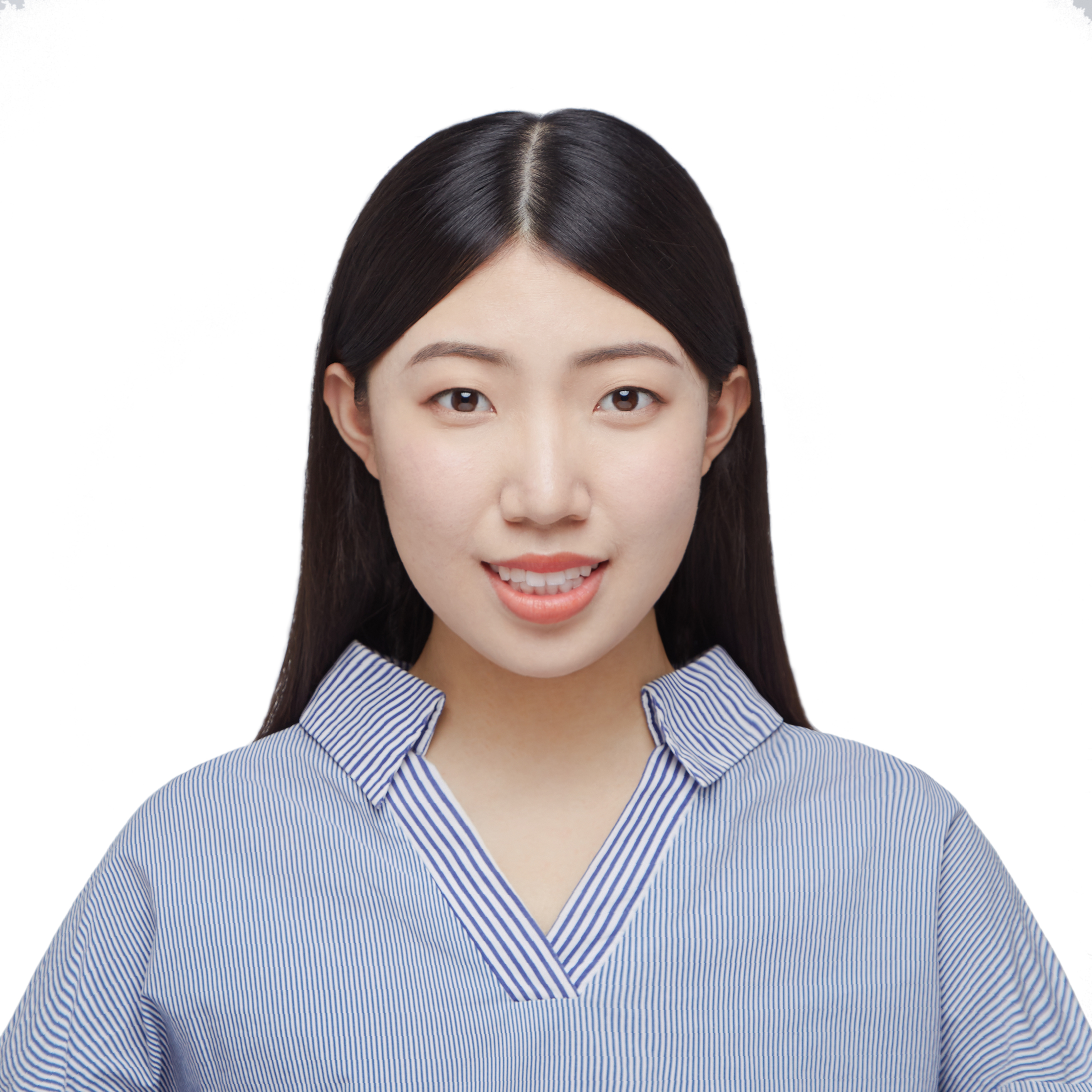}}]{Hanzhi Wang}
is currently a Ph.D. candidate at School of Information, Renmin University of China, advised by Professor Zhewei Wei.
She received her B.E. degree in Computer Science and Technology at School of Information, Renmin University of China in June 2019.
Her research focuses on the development of efficient graph analysis and learning algorithms. 
\end{IEEEbiography}

\begin{IEEEbiography}[{\includegraphics[width=1in,height=1.25in,clip,keepaspectratio]{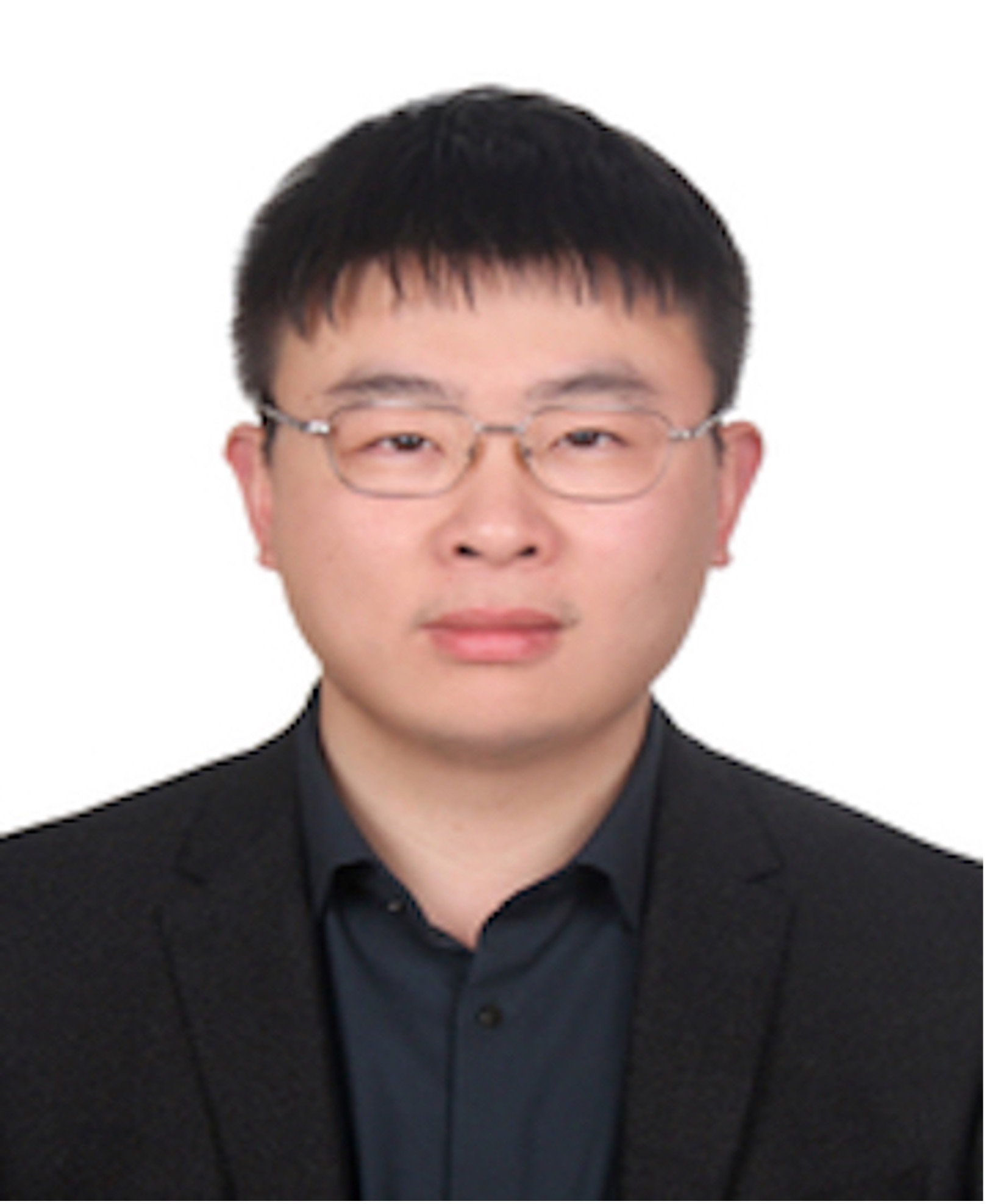}}]{Zhewei Wei}
is currently a Professor at Gaoling School of Artificial Intelligence, Renmin University of China.
He obtained his Ph.D. degree at Department of Computer Science and Engineering, HKUST in 2012.
He received the B.Sc. degree in the School of Mathematical Sciences at Peking University in 2008.
His research interests include graph algorithms, massive data algorithms, and streaming algorithms.
He was the Proceeding Chair of SIGMOD/PODS2020 and ICDT2021. 
\end{IEEEbiography}

\begin{IEEEbiography}[{\includegraphics[width=1in,height=1.25in,clip,keepaspectratio]{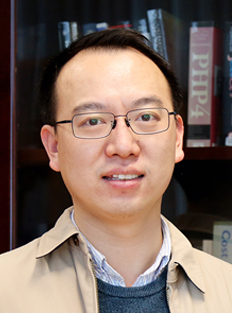}}]{Sibo Wang}
is an Assistant Professor in the Department of Systems Engineering and Engineering Management, Faculty of Engineering.
He received his B.E. in Software Engineering in 2011 from Fudan University and his Ph.D. in Computer Science in 2016 from Nanyang Technological University.
His main research area is database and data mining.
He is currently interested in graph data management, big data analysis, especially social network analysis, and efficient algorithms with indexing and approximation.
\end{IEEEbiography}

\begin{IEEEbiography}[{\includegraphics[width=1in,height=1.25in,clip,keepaspectratio]{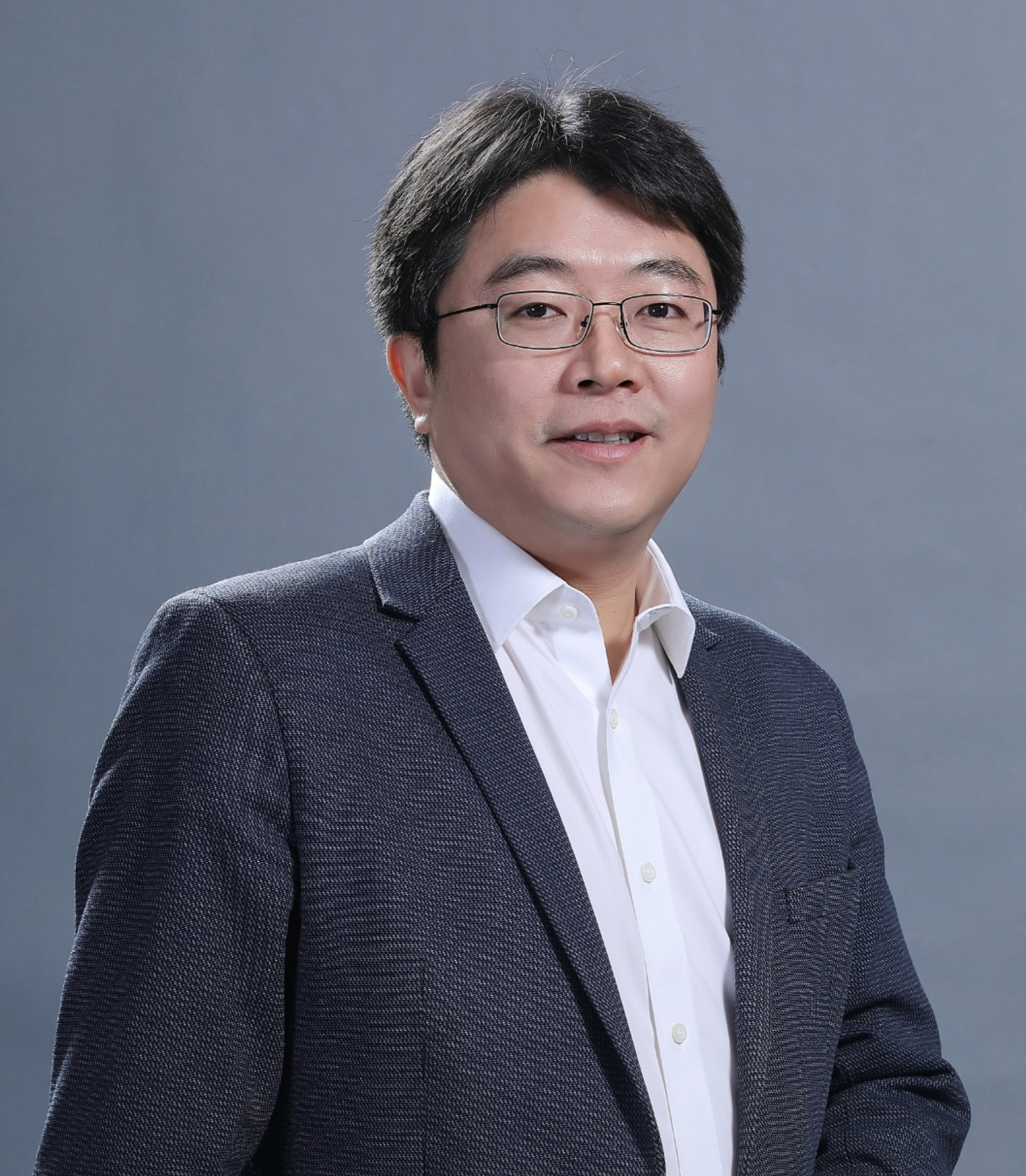}}]{Ji-Rong Wen}
is a professor, the dean of School of Information and the executive dean of Gaoling School of Artificial Intelligence at Renmin University of China.
His main research interests include information retrieval, data mining, and machine learning.
He once was a senior researcher and group manager of the Web Search and Mining Group at Microsoft Research Asia (MSRA).
He is the Associate Editor of ACM TOIS and IEEE TKDE, Honorary Chair of AIRS 2016, and PC Chair of SIGIR 2020.
He is also a chief scientist at Beijing Academy of Artificial Intelligence.
\end{IEEEbiography}

\end{document}